\documentclass[12pt,a4paper]{article}
\usepackage[T1]{fontenc}
\usepackage{fancyhdr}
\usepackage{amsmath,amsthm,amsfonts,amssymb}
\allowdisplaybreaks
\usepackage{epic,eepic}
\usepackage{graphicx}
\usepackage{authblk}
\usepackage{fullpage}
\usepackage[active]{srcltx}
\usepackage{enumitem}
\usepackage{booktabs}

\usepackage{framed,color}

\usepackage{float}%
\floatstyle{plaintop}%
\restylefloat{table}

\usepackage[font={small,it}]{caption}

\usepackage[table]{xcolor}

\usepackage[toc,page]{appendix}
\graphicspath{{Figures/}}

\usepackage[style=authoryear,backend=bibtex,maxcitenames=2,maxbibnames=99]{biblatex}
\bibliography{LossFunctions}

\usepackage{chngcntr}
\usepackage{bbm}

\usepackage{algpseudocode}
\usepackage{algorithm}
\usepackage{setspace}

\DeclareMathOperator*{\argmax}{arg\,max}
\DeclareMathOperator*{\argmin}{arg\,min}

\makeatletter
\AtEveryBibitem{%
  \global\undef\bbx@lasthash%
  \clearfield{}}
\makeatother

\theoremstyle{plain}

\theoremstyle{definition}

\theoremstyle{remark}

\numberwithin{equation}{section}
\numberwithin{figure}{section}

\date{\today}
\title{\textbf{Optimal Bayesian estimators for latent variable cluster models}}
\author[1,2,*]{Riccardo Rastelli}
\author[1,2]{Nial Friel}
\affil[1]{\footnotesize School of Mathematics and Statistics, University College Dublin, Ireland;}
\affil[2]{\footnotesize Insight: Centre for Data Analytics, Ireland.}

\begin{document}
\rowcolors{2}{gray!25}{white}
\counterwithout{figure}{section}
\counterwithout{figure}{subsection}
\counterwithout{equation}{section}
\counterwithout{equation}{subsection}

\maketitle
\begin{abstract}
\noindent In cluster analysis interest lies in probabilistically capturing partitions of individuals, items or observations into groups, such that those belonging to the same group share similar attributes or relational profiles.
Bayesian posterior samples for the latent allocation variables can be effectively obtained in a wide range of clustering models, including finite mixtures, infinite mixtures, hidden Markov models and block models for networks.
However, due to the categorical nature of the clustering variables and the lack of scalable algorithms, summary tools that can interpret such samples are not available.
We adopt a Bayesian decision theoretic approach to define an optimality criterion for clusterings, and propose a fast and context-independent greedy algorithm to find the best allocations.
One important facet of our approach is that the optimal number of groups is automatically selected, thereby solving the clustering and the model-choice problems at the same time.
We consider several loss functions to compare partitions, and show that our approach can accommodate a wide range of cases. 
Finally, we illustrate our approach on a variety of real-data applications for three different clustering models: Gaussian finite mixtures, stochastic block models and latent block models for networks.
\\

\end{abstract}

\baselineskip=17pt
\section{Introduction}
Cluster analysis plays a central role in statistics and machine learning, yet it is not immediately clear how one can appropriately summarise the output of partitions from a Bayesian clustering model.
This article seeks to address this impasse, proposing an optimality criterion for clusterings derived from decision theory, and a greedy algorithm to estimate the optimal partition and number of groups.
Clustering models are often represented as discrete latent variable models: each of the data objects corresponds to the elements of $\mathcal{V}=\left\{ 1,2,\dots,N\right\}$ and is characterised by a categorical latent variable 
$z=\left\{ 1,2,\dots,K\right\}$ denoting its group label.
Such variables are often called \textit{clustering variables} or \textit{allocations}.
Notable examples of latent variable clustering models include: 
product partition models \parencite{hartigan1990partition,barry1992product}, 
finite mixtures \parencite{mclachlan2004finite},
infinite mixtures (\textcite{quintana2006predictive} and references therein), 
latent block models for networks \parencite{nowicki2001estimation,govaert1995simultaneous}, 
hidden Markov models \parencite{macdonald1997hidden}.

The motivation for this paper ensues from the introduction within the statistical community of the so-called trans-dimensional samplers. 
One well known and widely used sampler is the \textit{reversible jump algorithm} of \textcite{green1995reversible}, 
extended to the context of finite mixtures by \textcite{richardson1997bayesian} and to hidden Markov models by \textcite{robert2000bayesian}.
Reversible jump Markov chain Monte Carlo allows one to explore a number of models with a single Markov chain that ``jumps'' between them, thereby estimating both the model parameters and the posterior model probabilities.
A more recent trans-dimensional Markov Chain Monte Carlo algorithm is the \textit{allocation sampler} introduced by \textcite{nobile2007bayesian}.
This takes advantage of the fact that, in some finite mixture models, the marginal posterior distribution of the allocation variables can be obtained by analytically integrating out all of the model parameters.
This allows one to use a \textit{collapsed} Gibbs sampler and obtain a posterior marginal sample for the clustering variables. 
One advantage of this method is that the number of groups can be inferred at each step from the clustering variables automatically, hence obtaining posterior probabilities for the different models.
The core idea of the allocation sampler has been recently extended to a number of frameworks, including latent class analysis \parencite{white2016bayesian}, latent block models \parencite{wyse2012block}, 
stochastic block models \parencite{mcdaid2013improved}, latent position models \parencite{friel2013bayesian}, and change point analysis \parencite{benson2016adaptive}.
In Bayesian nonparametrics a similar approach has been proposed by \parencite{neal2000markov,favaro2013mcmc} for Dirichlet process mixture models.

Both reversible jump and allocation sampler return a trans-dimensional sample for the allocations. 
Theoretically, such a sample contains all of the posterior information needed for the clustering of the data, however, interpreting such information is a very challenging task.
Since the allocations are categorical variables, usual summary statistics such as the mean, median and quantiles are not well defined. 
In addition, these Markov chain Monte Carlo algorithms are sensitive to label-switching issues \parencite{stephens2000dealing}, 
in fact, when using the latent variable representation, all mixture models are non-identifiable up to a permutation of the cluster labels.
In addition, the sample itself may be computationally impractical to handle, since even basic operations may require a cost that grows with $N^2$ or the square of the size of the sample.

The problem described really boils down to a very simple research question: we want to summarise the information provided by a sample of partitions into an optimal partition.
This issue has been addressed in several previous works, such as \textcite{strehl2003cluster,gionis2007clustering,dahl2009modal,fritsch2009improved}, 
where the authors propose a number of approaches that define a theoretical optimal partition and introduce algorithms to find it.
One critique to these contributions is that the proposed methodologies lack a sound theoretical background and they may be seen as ad-hoc.

In this work we use a Bayesian decision theoretic framework to define an optimality criterion for partitions, 
as previously proposed by \textcite{binder1978bayesian,lau2007bayesian,wade2015bayesian}.
From the Bayesian theoretical point of view, our approach defines the best possible solution to the partitioning problem using the information contained in the sample. 
Also, an important facet of this methodology is that it builds upon recent adaptations of the allocation sampler \parencite{wyse2012block,mcdaid2013improved,friel2013bayesian,white2016bayesian},
making up for one important shortcoming of these samplers: the interpretation of the results.

The essence of the decision theoretic framework lies in the definition of a loss function in the space of partitions, which is often a metric measuring how different two partitions are.
Then, the optimal partition is estimated as the one minimising the average loss with respect to the sample given. 
In the Bayesian perspective, this is equivalent to adopting a Bayes estimator (or \textit{Bayes action}), which is the decision minimising the \textit{Expected Posterior Loss} (EPL).

We propose a greedy algorithm as means to find the optimal partition, focusing on its computational complexity and scalability.
The algorithm can deal with a wide family of loss functions and requires only the sample of partitions as input. 
Hence our methodology has wide applicability and is the only scalable procedure that can be used to perform Bayesian clustering for a relatively arbitrary loss function.
One important advantage of our algorithmic frameworks is that the resulting optimal clustering automatically determines the optimal number of groups.

Previous works \parencite{lau2007bayesian,wade2015bayesian} were confined to the case of Bayesian nonparametric models. 
Here we stress that this approach is automatically extended to a very general clustering context, and hence we propose applications to several different frameworks.

The plan of the paper is summarised as follows: 
Section \ref{BayesianClusteringTheory} describes the theoretical foundations of Bayesian clustering;
in Section \ref{ChoiceOfTheLossFunction} we describe the properties of several loss functions to compare partitions, and we characterise the wide breadth to which our method extends;
in Section \ref{MinimisationOfTheExpectedPosteriorLoss} we introduce our greedy algorithm and analyse its complexity and features, whereas
Section \ref{ClassesOfEquivalencesWithinThePosteriorSample} shows an interesting procedure that can be used to potentially save an amount of computational time.
Finally, three applications to real datasets are proposed in Section \ref{RealDataExamples}: 
the galaxies' dataset for univariate Gaussian finite mixtures, 
the French political blogosphere for stochastic block models,
and the congressional voting data for latent block models. Section \ref{Conclusions} closes the paper with some final comments.

\section{Bayesian clustering: the theory}\label{BayesianClusteringTheory}
Let $\textbf{Z}$ be a $T\times N$ matrix, where, for every $t=1,\dots,T$ and $i=1,\dots,N$, $z_{ti}$ is a categorical variable 
(typically $z_{ti}\in\left\{1,2,\dots,N\right\}$) indicating the cluster label of observation $i$ at iteration $t$. 
The rows of $\textbf{Z}$ determine a sample of partitions of the same set $\mathcal{V} = \left\{1,2,\dots,N\right\}$, 
and we assume that such sample is drawn from the posterior distribution of a clustering model, given the observed data $\mathcal{Y}$.
An alternative representation of the sample would be $\left\{\textbf{z}^{(1)},\dots,\textbf{z}^{(T)}\right\}$, 
where $\textbf{z}^{(t)} = \left\{ z_{t1}, \dots, z_{tN}\right\}\in\mathcal{Z}$ corresponds to the $t$-th row of $\textbf{Z}$, and $\mathcal{Z}$ is the space of all partitions of $\mathcal{V}$.

Interest lies in conveying the information provided by the posterior sample into a single optimal partition. 
Bayesian decision theory offers an elegant approach to tackle this task, essentially recasting the clustering problem into one of decision making.

The first step consists of choosing a loss function $\mathcal{L}: \mathcal{Z}\times\mathcal{Z} \rightarrow \mathbb{R}$.
For any two partitions (hereafter also called \textit{decisions}) $\textbf{a}$ and $\textbf{z}$, 
the quantity $\mathcal{L}\left( \textbf{a},\textbf{z} \right)$ indicates the loss occurring when the decision $\textbf{a}$ is chosen while $\textbf{z}$ is the correct partition.
The choice of the loss function adopted is completely arbitrary and supposedly situational, nonetheless some loss functions have interesting features and tend to work well in 
many contexts. 
A loss function is not necessarily a distance in the space of partitions although this is often regarded as a desirable property, since it helps particularly in the 
interpretation and representation of the results.

An optimal decision (also called \textit{Bayes action}) is one minimising the \textit{expected posterior loss}, defined as:
\begin{equation}\label{epl1}
 \Psi\left( \textbf{a} \right) := \mathbb{E}_\textbf{z}\left[\mathcal{L}\left( \textbf{a},\textbf{z} \right)\middle\vert \mathcal{Y}\right] 
 = \sum_{\textbf{z}\in\mathcal{Z}} \pi\left( \textbf{z}\middle\vert \mathcal{Y}\right) \mathcal{L}\left( \textbf{a},\textbf{z} \right).
\end{equation}
Considering that the posterior sample $\left\{\textbf{z}^{(1)},\dots,\textbf{z}^{(T)}\right\} \sim \pi\left(\ \cdot\ \middle\vert \mathcal{Y}\right)$ is available,
for every decision $\textbf{a}\in\mathcal{Z}$, an unbiased estimator of the associated expected posterior loss results as:
\begin{equation}\label{epl2}
 \psi\left( \textbf{a} \right) = \frac{1}{T}\sum_{t=1}^T \mathcal{L}\left( \textbf{a},\textbf{z}^{(t)} \right) \approx \Psi\left( \textbf{a} \right).
\end{equation}

We aim then at finding the decision $\hat{\textbf{a}}$ minimising the approximate expected posterior loss:
\begin{equation}\label{epl3}
 \hat{\textbf{a}} = \displaystyle\argmin_{\substack{\textbf{a}\in\mathcal{Z}}} \psi\left( \textbf{a} \right).
\end{equation}

\section{Choice of the loss function}\label{ChoiceOfTheLossFunction}
\subsection{Common loss functions}
Given the sample $\textbf{Z}$, a naive but fast method to obtain an optimal clustering would be to consider 
the single partition that obtained the highest posterior value during the sampling, i.e.:
\begin{equation}\label{map1}
 \hat{\textbf{a}}_{MAP} = \displaystyle\argmax_{\substack{t=1,2,\dots,T}} \pi\left( \textbf{z}^{(t)}\middle\vert\mathcal{Y} \right).
\end{equation}
In a decision theoretic context, this is equivalent to choosing a 0$-$1 loss defined as:
\begin{equation}\label{01loss}
\rowcolors{1}{}{}
 \mathcal{L}\left( \textbf{a},\textbf{z} \right)= \begin{cases}
                                                   1&\mbox{ if }\textbf{a}\not\equiv\textbf{z},\\
                                                   0&\mbox{ if }\textbf{a}\equiv\textbf{z};
                                                  \end{cases}
\end{equation}
since the Bayes action minimising \eqref{epl3} would simply be the mode of the sample.
The sign ``$\equiv$'' here means that there exists a label permutation $\sigma$ such that $\sigma\left( a_i \right) = z_i$, $\forall i\in\mathcal{V}$.
Reading the definition in \eqref{01loss}, the loss is zero iff the partitions are equivalent. 
In all of the other cases, the loss is $1$ regardless of how different the partitions actually are.
This peculiar behaviour makes the 0$-$1 loss rather unappealing as means to compare partitions.

Note that all of the clustering algorithms that return a MAP estimate can be interpreted in this context as tools minimising the expected 0$-$1 loss, 
although they normally do not require the sample $\textbf{Z}$, and are computationally cheap.
Hence MAP estimates may be criticised since in the Bayesian paradigm the corresponding loss is not particularly sensible.

Another loss function that is commonly used is the quadratic loss, which gives the posterior mean as Bayes action. 
However, in a clustering context this has little meaning due to the categorical nature of the variables, which makes any sort of averaging of allocations not particularly meaningful.

\subsection{Loss functions to compare partitions}
A more sensible approach would be to choose a loss function that is specifically designed to compare partitions. 
In recent years, many measures to compare partitions have been proposed, each with very different properties and characteristics.
The works of \textcite{meilua2007comparing,vinh2010information,wade2015bayesian} and references therein offer an excellent overview.

A common approach used to compare partitions (here $\textbf{a}$ and $\textbf{z}$ denote two arbitrary partitions with $K_{\textbf{a}}$ and $K_{\textbf{z}}$ groups, respectively) relies on the $K_{\textbf{a}} \times K_{\textbf{z}}$ contingency matrix
(or confusion matrix), whose entries are defined as:
\begin{equation}\label{contingency1}
 n^{\textbf{a},\textbf{z}}_{gh} = \sum_{i=1}^{N} \mathbbm{1}_{\left\{ a_i=g\right\}}\mathbbm{1}_{\left\{ z_i=h\right\}}
\end{equation}
where $g$ varies among the groups of $\textbf{a}$, and $h$ among those of $\textbf{z}$. 
The entries of such a matrix simply count the number of items that $\textbf{a}$ classifies in group $g$ and $\textbf{z}$ classifies in group $h$, for every $g$ and $h$.

Here, we focus on loss functions that depend on $\textbf{a}$ and $\textbf{z}$ only through the entries of $\textbf{n}^{\textbf{a},\textbf{z}}$.
This is a fairly general and reasonable assumption which is in line with the theory developed by \textcite{binder1978bayesian}; 
in fact, most metrics can be transformed into functions of the counts (see \textcite{vinh2009information} and references therein).

We assume that the loss function has the following representation:
\begin{equation}\label{generalloss1}
 \mathcal{L}\left( \left\{n^{\textbf{a},\textbf{z}}_{gh}\right\}_{g,h}, \left\{n^{\textbf{a}}_{g}\right\}_{g}, \left\{n^{\textbf{z}}_{h}\right\}_{h} \right) = 
 f_0\left( \sum_{g=1}^{K_{\textbf{a}}}\sum_{h=1}^{K_{\textbf{z}}} f_1\left( n^{\textbf{a},\textbf{z}}_{gh} \right), \sum_{g=1}^{K_{\textbf{a}}} f_2\left( n^{\textbf{a}}_{g} \right), \sum_{h=1}^{K_{\textbf{z}}} f_3\left( n^{\textbf{z}}_{h} \right) \right)
\end{equation}
where $f_0,\ f_1,\ f_2,\ f_3$ are real valued functions that can be evaluated in constant time and $n^{\textbf{a}}_{g}$ and $n^{\textbf{z}}_{h}$ indicate the sizes of group $g$ and $h$, respectively, i.e.:
\begin{equation}\label{contingency2}
 n^{\textbf{a}}_{g} = \sum_{h=1}^{K_{\textbf{z}}} n^{\textbf{a},\textbf{z}}_{gh}, 
 \hspace{2cm} n^{\textbf{z}}_{h} = \sum_{g=1}^{K_{\textbf{a}}} n^{\textbf{a},\textbf{z}}_{gh}.
\end{equation}
for every $g=1,\dots,K_{\textbf{a}}$ and $h=1,\dots,K_{\textbf{z}}$.
The assumption determined by \eqref{generalloss1} is actually not restrictive: most of the commonly used loss functions for partitions satisfy this condition. 
We note that the arguments of the function $f_0$ include the following quantities as special cases:
\begin{itemize}
 \item The entropies of $\textbf{a}$ and $\textbf{z}$, describing the uncertainty associated to $\textbf{a}$ and $\textbf{z}$, respectively:
 \begin{equation}
  H\left( \textbf{a} \right) = -\sum_{g=1}^{K_{\textbf{a}}} \frac{n^{\textbf{a}}_{g}}{N} \log_2 \frac{n^{\textbf{a}}_{g}}{N}; \hspace{2cm}H\left( \textbf{z} \right) = -\sum_{h=1}^{K_{\textbf{z}}} \frac{n^{\textbf{z}}_{h}}{N}
  \log_2\frac{n^{\textbf{z}}_{h}}{N}.
 \end{equation}
 \item The joint entropy of $\textbf{a}$ and $\textbf{z}$:
 \begin{equation}
  H\left( \textbf{a},\textbf{z} \right) = -\sum_{g=1}^{K_{\textbf{a}}}\sum_{h=1}^{K_{\textbf{z}}} \frac{n^{\textbf{a},\textbf{z}}_{gh}}{N}\log_2\frac{n^{\textbf{a},\textbf{z}}_{gh}}{N}.
 \end{equation}
 This describes instead the uncertainty of the random variable with pdf given by the quantities $n^{\textbf{a},\textbf{z}}_{gh}/N$, for every $g$ and $h$.
 \item The mutual information, which can be evaluated from the entropies and joint entropy:
 \begin{equation}\label{mutual1}
  I\left( \textbf{a},\textbf{z} \right) = H\left( \textbf{a} \right) + H\left( \textbf{z} \right) - H\left( \textbf{a},\textbf{z} \right).
 \end{equation}
 This quantity is particularly meaningful and has been advocated in a normalised version by \textcite{strehl2003cluster} as a distance measure between partitions.
\end{itemize}
Note the common convention that $x\log_2 x = 0$ if $x=0$.
Evidently these information-based quantities can be obtained as special cases of the functions $f_1$, $f_2$ and $f_3$, making our assumption rather general and broadly satisfied.\\

Here follows a brief description of some well-known loss functions that can be considered with our approach.
\paragraph{Binder's loss (B).} We use a special case of a more general formula first introduced by \textcite{binder1978bayesian}:
 \begin{equation}\label{binder1}
  \mathcal{L}_{B}\left( \textbf{a},\textbf{z} \right) = 
  \frac{1}{2}\sum_{g=1}^{K_{\textbf{a}}} \left(n^{\textbf{a}}_g\right)^2 + \frac{1}{2}\sum_{h=1}^{K_{\textbf{z}}} \left(n^{\textbf{z}}_h\right)^2 - 
  \sum_{g=1}^{K_{\textbf{a}}}\sum_{h=1}^{K_{\textbf{z}}} \left(n^{\textbf{a},\textbf{z}}_{gh}\right)^2.
 \end{equation}
This loss is equivalent to the Hamming distance \parencite{meilua2012local} and to the Rand index \parencite{rand1971objective}.
Binder's loss has an interesting property that simplifies greatly the minimisation of \eqref{epl3}.
One can in fact easily construct a so-called posterior similarity matrix of size $N\times N$, whose entries $b_{ij}$ denote the estimated posterior probability of $i$ and $j$ being allocated to the same group, 
for every $i$ and $j$ in $\mathcal{V}$.
Then, the Binder Bayes action satisfies:
\begin{equation}\label{binder2}
 \hat{\textbf{a}}_B = \displaystyle\argmin_{\substack{\textbf{a}\in\mathcal{Z}}}\sum_{i<j}\left[ \mathbbm{1}_{\left\{ a_i = a_j\right\}} - b_{ij}\right]
\end{equation}
where $\mathbbm{1}_\mathcal{A}$ is equal to $1$ if the event $\mathcal{A}$ is true or zero otherwise.
This simplifies the minimisation problem since \eqref{binder2} depends on the sample only through the posterior similarity matrix, which can be effectively computed beforehand.

\paragraph{The variation of information (VI).} This loss is one we particularly focus on in this paper, and is defined as:
\begin{equation}\label{viloss1}
 \mathcal{L}_{VI}\left( \textbf{a},\textbf{z} \right) = 2H\left( \textbf{a},\textbf{z} \right) - H\left( \textbf{a} \right) - H\left( \textbf{z} \right).
\end{equation}
The VI loss, first studied in \textcite{meilua2007comparing}, has received an increasing amount of attention in the last decade, 
mainly due to its strong mathematical foundations and practical efficiency. In the paper by \textcite{meilua2007comparing} as well as in subsequent 
works such as \textcite{wade2015bayesian}, the mathematical properties and behaviour of the VI loss have been deeply studied. 
We mention that this loss is a metric, that it forms a lattice and that it is horizontally and vertically aligned in the space of partitions.
In addition, it is invariant to label-switching, i.e. switching labels for either $\textbf{a}$ or $\textbf{z}$ will not affect the value $\mathcal{L}_{VI}\left( \textbf{a},\textbf{z} \right)$.
More details regarding the theoretical properties of the VI loss can be found in \textcite{meilua2007comparing}.

\paragraph{The normalised variation of information (NVI).} This loss is defined as:
 \begin{equation}
  \mathcal{L}_{NVI}\left( \textbf{a},\textbf{z} \right) = 1 - \frac{I\left( \textbf{a},\textbf{z} \right)}{H\left( \textbf{a},\textbf{z} \right)}.
 \end{equation}
 The normalised version of the VI loss takes values in $[0,1]$. This scale-invariance may facilitate the interpretation and the comparisons of partitions under different conditions.
 Since we adopt an optimisation approach, this feature is not crucial in our framework due to the partitions always referring to the same set of individuals. 

\paragraph{The normalised information distance (NID).} This loss is defined as:
 \begin{equation}
  \mathcal{L}_{NID}\left( \textbf{a},\textbf{z} \right) = 1-\frac{I\left( \textbf{a},\textbf{z} \right)}{\max\left\{H\left( \textbf{a} \right),H\left( \textbf{z} \right)\right\}}.
 \end{equation}
The NID loss has been advocated in \textcite{vinh2010information} as a general purpose - context independent - loss function with desirable behaviours.

\section{Minimisation of the expected posterior loss}\label{MinimisationOfTheExpectedPosteriorLoss}
An exhaustive search within $\mathcal{Z}$ becomes impractical even for very small $N$ (the cardinality of $\mathcal{Z}$ is a number with more than $100$ digits if $N=100$).
Therefore, the minimisation can be seen as a binary programming optimisation problem which is known to be NP-hard, and hence not solvable through exact methods.

Also, the objective function requires the calculation of the sum in \eqref{epl2} at each evaluation. 
Getting a new posterior sample at each step is not a practical option, hence the same sample is used for all of the evaluations of \eqref{epl2}. 
Nonetheless, even a single evaluation of the objective function can become computationally burdensome when the size of the sample is large.
Therefore, the decision theoretic approach becomes soon impractical as $N$ and $T$ increase, and finding scalable procedures is crucial.
In this section we introduce a new algorithm that, using greedy updates, is able to estimate the Bayes action for the wide family of loss functions satisfying \eqref{generalloss1}, requiring in input only the posterior sample of partitions.

\subsection{Greedy algorithm}
Heuristic greedy algorithms have been recently rediscovered as a means to maximise the so-called \textit{exact Integrated Complete Likelihood} in various contexts: 
stochastic block models \parencite{come2015model}, latent block models \parencite{wyse2014inferring}, Gaussian finite mixtures \parencite{bertoletti2015choosing}. 
Similar approaches have also been proposed in Bayesian nonparametrics for Dirichlet prior mixtures \parencite{raykov2014simple} although in this case they did not 
cast the clustering problem into the optimisation of an exact model-based clustering criterion.
Among the many papers adopting types of greedy optimisation, we find the approaches of \textcite{besag1986statistical,strehl2003cluster,newman2004fast} particularly related to ours.

We propose a greedy algorithm that updates a partition by changing the cluster memberships of single observations using a greedy heuristic, hence decreasing the 
expected posterior loss of the partition at each step. As input, the algorithm only requires a starting partition, the posterior sample $\textbf{Z}$ and a user-specified 
parameter $K_{up}$, equal to the maximum number of groups allowed (a reasonable default value would be $K_{up}=N$). The algorithm cycles over the observations in random 
order, and, for each of these, it tries all of the possible reallocations, eventually choosing the one giving the best decrease in the objective function.
The notation $\textbf{a}_{i:r\rightarrow s}$ denotes the partition $\textbf{a}$ where the observation $i$ has been reallocated from group $r$ to $s$.
At each move, the number of groups may increase (if the observation is reallocated to an empty group) or decrease (if a group is left empty), although the latter scenario is much more frequent. 
Due to the low probability of creating new groups, it is generally advisable to start with a partition made of close to $K_{up}$ groups.
The procedure stops when a complete sweep over all observations yields no change in the expected posterior loss.
The pseudo-code for the algorithm is shown in Algorithm \ref{algorithm1}.
\begin{algorithm}[htb]
\begin{spacing}{1.2}
\begin{algorithmic}[1]
\State Let $\textbf{a}$ be the starting partition.
\State Set $\psi_{\textbf{a}} = \psi\left( \textbf{a} \right)$.
\State Set STOP to $false$.
\While{STOP is $false$}
\State $\psi_{stop} = \psi_{\textbf{a}}$.
\State Set $\mathcal{V} = \left\{1,2,\dots,N \right\}$.
\While{$\mathcal{V}$ is not empty}
\State Pick $i$ at random from $\mathcal{V}$ and delete it from $\mathcal{V}$.
\State For every $s=1,\dots,K_{up}$, evaluate $\psi\left( \textbf{a}_{i:a_i\rightarrow s} \right)$.
\State Move $i$ to $\hat{s}=\displaystyle\argmax_{\substack{s=1,\dots,K_{up}}} \psi\left( \textbf{a}_{i:a_i\rightarrow s} \right)$.
\State Update $\psi_{\textbf{a}} = \psi\left( \textbf{a} \right)$.
\EndWhile
\If {$\psi_{stop} = \psi_{\textbf{a}}$} \State STOP = $true$. \EndIf
\EndWhile
\State Return $\textbf{a}$ and $\psi_{\textbf{a}}$.
\end{algorithmic}
\caption{Greedy algorithm}
\label{algorithm1}
\end{spacing}
\end{algorithm}

Due to the greedy nature of this procedure, the algorithm is bound to return a local optimum, rather than a global one. Consequently, several restarts with different 
initial partitions may be required. However, convergence is usually reached in very few iterations, in each run. 
Regarding the starting partition, this may either be chosen at random or it may be set to be the clustering yielding the highest posterior value as in \eqref{map1}. 
A possible alternative lies in between the two cases, i.e. the MAP partition may be changed to some extent by reallocating some observations at random.

One interesting feature of the greedy algorithm is that the whole space of partitions is explored, hence the optimal partitions may differ substantially from all of the 
clusterings in the sample. In fact, many non-optimal solutions may have higher posterior values than the optimal one.
In contrast to \textcite{come2015model,wyse2014inferring}, we do not perform any final merge step, as in most cases this did not improve the results.

\subsection{Complexity}
The basic operation that determines the complexity of the greedy optimisation is the evaluation of the variation in the objective function when a possible reallocation is tested
(line $9$ in the pseudo-code \ref{algorithm1}). Assume that the move from $\textbf{a}$ to $\textbf{a}_{i:r\rightarrow s}$ is being tested, for some groups $r$ and $s$.
The following quantity needs to be evaluated:
\begin{equation}
 \Delta \psi := \psi \left( \textbf{a}_{i:r\rightarrow s} \right) - \psi \left( \textbf{a} \right) = \frac{1}{T}\sum_{t=1}^{T} \left[ \mathcal{L}\left( \textbf{a}_{i:r\rightarrow s},\textbf{z}^{(t)} \right) - \mathcal{L}\left( \textbf{a},\textbf{z}^{(t)} \right) \right],
\end{equation}
which in turn requires, $\forall t= 1,\dots, T$:
\begin{equation}
 \Delta \mathcal{L}^{(t)} := \mathcal{L}\left( \textbf{a}_{i:r\rightarrow s},\textbf{z}^{(t)} \right) - \mathcal{L}\left( \textbf{a},\textbf{z}^{(t)} \right).
\end{equation}
For a certain $t$, the move only affects two entries of $\textbf{n}^{\textbf{a}}$ (i.e. $n^{\textbf{a}}_{r}$ and $n^{\textbf{a}}_{s}$) and two entries of $\textbf{n}^{\textbf{a},\textbf{z}^{(t)}}$ 
(i.e. $n^{\textbf{a},\textbf{z}^{(t)}}_{rv}$ and $n^{\textbf{a},\textbf{z}^{(t)}}_{sv}$, where $v = z_{ti}$).
This means that the change in the arguments of $f_0$ can be evaluated in a constant time, hence making the cost of evaluating $\Delta \psi \sim \mathcal{O}\left( T \right)$.

Since the algorithm tries all possible moves for each observation, the overall computational cost is $\mathcal{O}\left( TNK_{up} \right)$.

\subsection{Comparisons with other algorithms}
Both \textcite{lau2007bayesian} and \textcite{wade2015bayesian} propose original algorithmic frameworks to minimise an expected posterior loss. 
While \textcite{lau2007bayesian} only focus on Binder's loss, \textcite{wade2015bayesian} also extend the procedure to the VI loss, albeit resorting to an approximation of the objective function.
Both methodologies take advantage of the posterior similarity matrix representation, briefly pointed out in \eqref{binder2}.
Note that this representation is exclusive to the Binder's loss, hence these approaches lack the possibility to be generalised to other loss functions, unless approximations are introduced.

The computational cost for an evaluation of the objective function \eqref{binder2} does not depend on $T$, since the posterior information contained in the sample is summarised in the posterior similarity matrix.
The calculation of the posterior similarity matrix itself requires $\mathcal{O}\left( TN^2 \right)$ operations, yet this can be performed offline and it is unlikely to impact the overall computing time.

On the other hand, our algorithm does not require a $N^2$ cost at any stage, hence it should be preferrable when the number of observations to classify is very large.
We note that, due to the dependence of the complexity on $T$, our algorithm will benefit if the sample is small and thinned with a large lag.
A trade-off between the reliability of the posterior sample and computing time should be assessed, in that one should provide a sample that is as small as possible but not so small that
the approximation to the posterior distribution is not reliable.
As concerns the dependency on $K_{up}$, ideally one should choose $K_{up}=N$, but this evidently would make the procedure impractical in large $N$ scenarios. 

More generally, the computational cost of the algorithm may be compared to the complexity of the sampler used to get the posterior sample. 
In fact, one key advantage of the collapsed Gibbs samplers proposed in \textcite{nobile2007bayesian,mcdaid2013improved,wyse2012block} is their computational efficiency.
The posterior sample returned by these samplers is necessary to perform the minimisation of the expected posterior loss. 
Hence, an ideal complexity for the optimisation problem should be not higher than that required by the sampler in the first place. 
Unfortunately, when analysing these samplers, new quantities (the number of dimensions for Gaussian finite mixtures, or the number of edges in block models) come into play, 
making a strict comparison of the complexity not possible. 
However, in our applications we noticed that for stochastic block models and latent block models the computational bottleneck was set by the samplers, and not by the greedy algorithm. 

\section{Classes of equivalences in the posterior sample}\label{ClassesOfEquivalencesWithinThePosteriorSample}
Since the sample space $\mathcal{Z}$ is discrete, the posterior sample $\textbf{Z}$ may contain repetitions, due to the sampler returning to the same partition during the sampling procedure.
This suggests that, regardless of the partition $\textbf{a}$, a number of the calculations required to obtain $\mathcal{L}\left( \textbf{a},\textbf{z} \right)$ is redundant. 
In fact, given a partition $\textbf{z}$, the following holds:
\begin{align}
 \mathcal{L}\left( \textbf{a},\textbf{z}^{(t)} \right) &= \mathcal{L}\left( \textbf{a},\textbf{z} \right);\\
 \mathcal{L}\left( \textbf{a}_{i\rightarrow g},\textbf{z}^{(t)} \right) &= \mathcal{L}\left( \textbf{a}_{i\rightarrow g},\textbf{z} \right).
\end{align}
for all $i=1,\dots,N$ and $g=1,\dots,K_{up}$ and $\forall t:\ \textbf{z}^{(t)} \equiv \textbf{z}$.

It follows that the posterior sample can be summarised into the sample of its unique rows $\tilde{\textbf{Z}} = \left\{\tilde{\textbf{z}}^{(1)},\dots,\tilde{\textbf{z}}^{(\tilde{T})}\right\}$
and a vector of counts $\boldsymbol{\omega} = \left\{ \omega^{(1)},\dots,\omega^{(\tilde{T})}\right\}$ describing how many times the corresponding partition appears in the original sample $\textbf{Z}$.
Therefore the approximate expected posterior loss can be equivalently written as:
\begin{equation}\label{epl4}
 \psi\left( \textbf{a} \right) = \frac{1}{\tilde{T}}\sum_{t=1}^{\tilde{T}} \omega^{(t)}\mathcal{L}\left( \textbf{a},\tilde{\textbf{z}}^{(t)} \right).
\end{equation}
A similar reasoning can be used to make the calculation of $\psi\left( \textbf{a}_{i\rightarrow g} \right)$ more efficient.

The main difficulty in applying the technique just described lies in identifying the new representation efficiently. One problem consists in the implementation of the operator ``$\equiv$'' since 
partitions should be compared up to a permutation of the labels. To solve this, we use a procedure described in \textcite{strehl2003cluster} that defines a unique labelling for all partitions:
the first item is assigned to cluster $1$, and then iteratively the next item is assigned either to an existing cluster or to the next empty cluster. 
Using this re-labelling, any two equivalent partitions will be transformed into the same sequence of digits in a computational time $\mathcal{O}\left( TN \right)$.

Furthermore, the same vector can be seen as a number in base$-K_{up}$ representation which uniquely identifies the corresponding partition and the equivalence class imposed by ``$\equiv$''.
Hence a sorting algorithm can be used to reorder the sample according to such identifiers, 
for a computational cost of $\mathcal{O}\left( NT\log T \right)$, where $N$ is the cost of a single comparison of partitions.
Once the partitions are sorted, the unique set and the corresponding weights can be obtained in $\mathcal{O}\left( TN \right)$.

The advantage provided by this representation heavily depends on the dataset and on the corresponding marginal posterior distribution: less repetitions will appear if the posterior is flat and the 
partitioning very uncertain. On the other hand, the computational savings may be substantial in cases where only few partitions have a high posterior value.

Note that the sorting procedure creates a new computational bottleneck in the case where $\log T > K_{up}$. 
However, we found this is not relevant in practical terms and negligible when compared to the computational time demanded by the actual optimisation.

In a machine learning context, the weighted sample $\tilde{\textbf{Z}}$ may be interpreted as a cluster ensemble problem, 
whereby each partition corresponds to the output of a clustering algorithm and the counts are weights describing the relative (possibly subjective) importance of the solution. 
Our methodology may be applied in this scenario without further modifications, providing a sound background to the decision making process.

\section{Real data examples}\label{RealDataExamples}
In this Section, we provide three applications of our methodology to different clustering contexts, and compare the results obtained with previous analyses.
To avoid confusion, we show the results only for the VI loss, and note that the other losses lead to similar partitions.

\subsection{Galaxies' dataset}
\subsubsection{The data}
The dataset considered is composed of the velocities of $82$ distant galaxies diverging from the Milky Way. 
Interest lies in understanding whether velocity can be used to discriminate clusters of galaxies. 
The dataset has been first analysed from a statistical point of view in \textcite{roeder1990density}, and has been re-proposed in numerous papers dealing with mixture models,
including \textcite{richardson1997bayesian,stephens2000dealing,wade2015bayesian}. 

\subsubsection{The model}
The observed data is denoted by $\mathcal{Y}=\left\{ y_1,\dots, y_N\right\}$, where $N=82$ and $y_i\in\mathbb{R}$ for every $i=1,\dots,N$.
As in \textcite{bertoletti2015choosing}, a Gaussian finite mixture model is adopted:
\begin{equation}
 p\left( \mathcal{Y}\middle\vert \boldsymbol{\lambda}, \boldsymbol{\mu},\textbf{r} \right) = \prod_{i=1}^{N}\sum_{g=1}^{K}\lambda_g\mathcal{N}\left( y_i;\ \mu_g,\ \frac{1}{r_g} \right);
\end{equation}
where $\lambda_1,\dots, \lambda_g$ are the mixture weights and $\mathcal{N}\left(\ \cdot\ ; \mu, \frac{1}{r} \right)$ denotes the univariate Gaussian distribution with mean $\mu$ and variance $1/r$.
The number $K$ of Gaussian components in the mixture is unknown and hence to be inferred.

Following a latent variable framework, an allocation variable $z_i$ is associated to each observation, denoting which Gaussian component has generated the corresponding $y_i$:
\begin{equation}
 p\left( y_i \middle\vert z_i = g, \boldsymbol{\mu}, \textbf{r}\right) = \mathcal{N}\left( y_i;\ \mu_g,\ \frac{1}{r_g} \right).
\end{equation}
This allows a more tractable expression for the likelihood, conditionally on the allocations:
\begin{equation}
 p\left( \mathcal{Y}\middle\vert \textbf{z},\boldsymbol{\mu},\textbf{r} \right) = \prod_{g=1}^{K} \prod_{i: z_i=g} \mathcal{N}\left( y_i;\ \mu_g,\ \frac{1}{r_g} \right).
\end{equation}

We specify a Bayesian hierarchical structure on both the likelihood parameters and the allocation variables.
For every group $g$, the parameters $r_g$ and $\mu_g$ are independent realisations of a Gamma$\left( \gamma,\delta \right)$ and a Gaussian$\left( 0, \left[\tau r_g\right]^{-1}\right)$, respectively.
As concerns the allocations, these are distributed as independent Multinomials with parameters $\boldsymbol{\theta}=(\theta_1,\dots,\theta_K)$, 
where $\boldsymbol{\theta}$ is a Dirichlet distributed random vector.
The hyperparameters are set as in \textcite{bertoletti2015choosing}: $\tau = 0.01$, $\gamma = 0.5$, $\delta = 0.5$, and Dirichlet hyperparameter $\alpha = 4$.

Since conjugate priors are used, most of the model parameters can be integrated out analytically. Hence the following marginal distributions can be obtained in exact form for the data and allocations:
\begin{align}
 p\left( \mathcal{Y}\middle\vert \textbf{z},\tau,\gamma,\delta \right) &= 
  \prod_{g=1}^{K} \int_0^\infty p\left( r_g\middle\vert\gamma,\delta \right)\int_{-\infty}^{+\infty} p\left( \mu_g\middle\vert \tau,r_g\right)\prod_{i: z_i = g}p\left( y_i \middle\vert z_i = g, \mu_g, r_g\right)d\mu_gdr_g; \\
 p\left( \textbf{z}\middle\vert \alpha \right) &= \int_\Theta p\left( \textbf{z}\middle\vert \boldsymbol{\theta} \right)p\left( \boldsymbol{\theta}\middle\vert\alpha \right)d\boldsymbol{\theta}.
\end{align}
More details on the integrations can be found in \textcite{nobile2007bayesian,bertoletti2015choosing}.
A consequence of these results is that the marginal posterior for the allocations can be obtained analytically, too:
\begin{equation}
 p\left( \textbf{z}\middle\vert \mathcal{Y} \right) \propto p\left( \mathcal{Y}\middle\vert \textbf{z},\tau,\gamma,\delta \right) p\left( \textbf{z}\middle\vert \alpha \right)
\end{equation}
Such marginal posterior distribution can be used as target in a Markov chain Monte Carlo sampler, thereby obtaining the posterior sample $\textbf{Z}$.
Note that, since all of the model parameters have been integrated out, trans-dimensional moves can be easily implemented, so that the chain effectively explores all of the possible models.
In Appendix \ref{ASimplifiedAllocationSampler}, a general algorithm to sample from this distribution is described. The same sampler is used to get the posterior sample $\textbf{Z}$ for the galaxies' dataset.

\subsubsection{Results}
\begin{figure}[htb]
\centering
\includegraphics[width=0.49\textwidth]{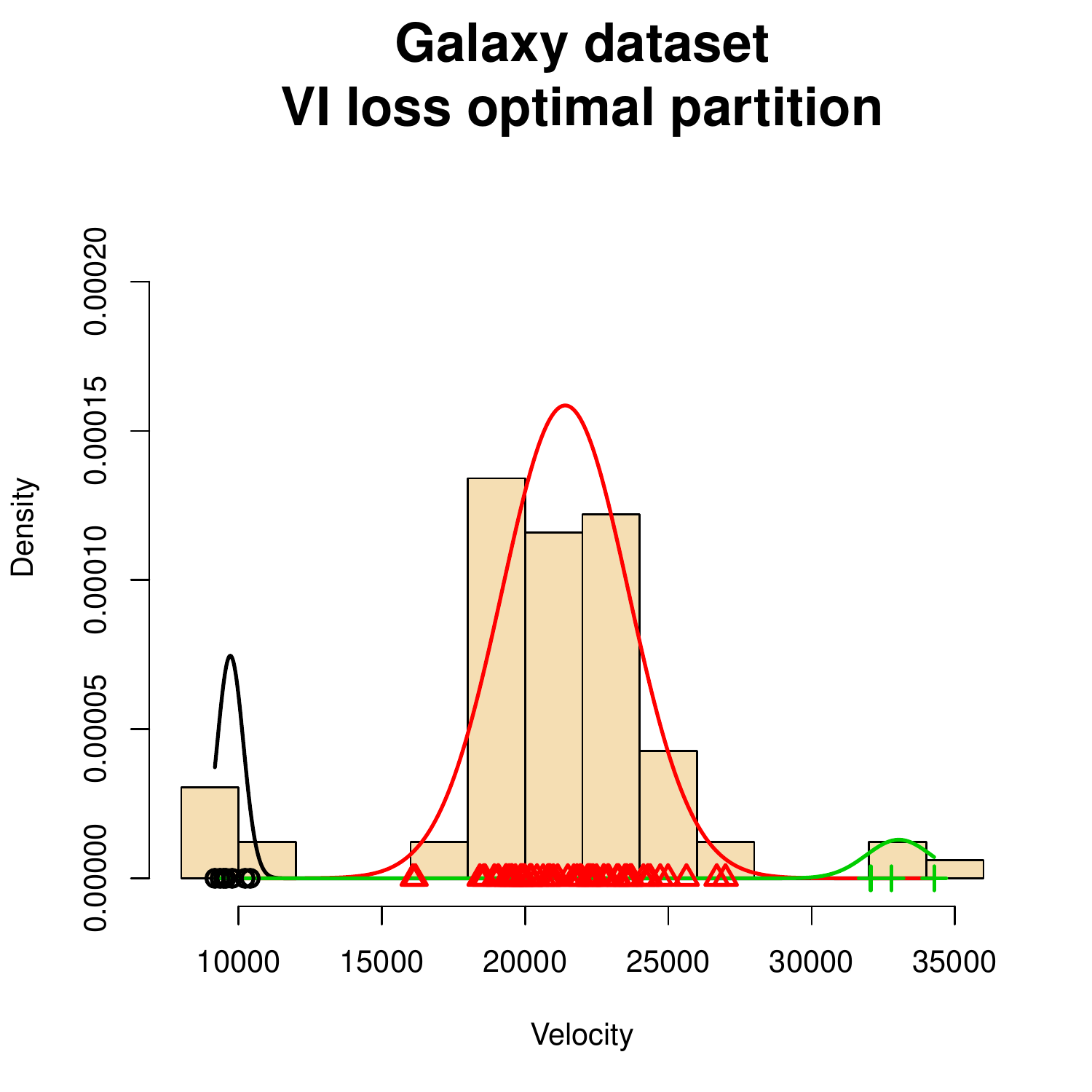}
\includegraphics[width=0.49\textwidth]{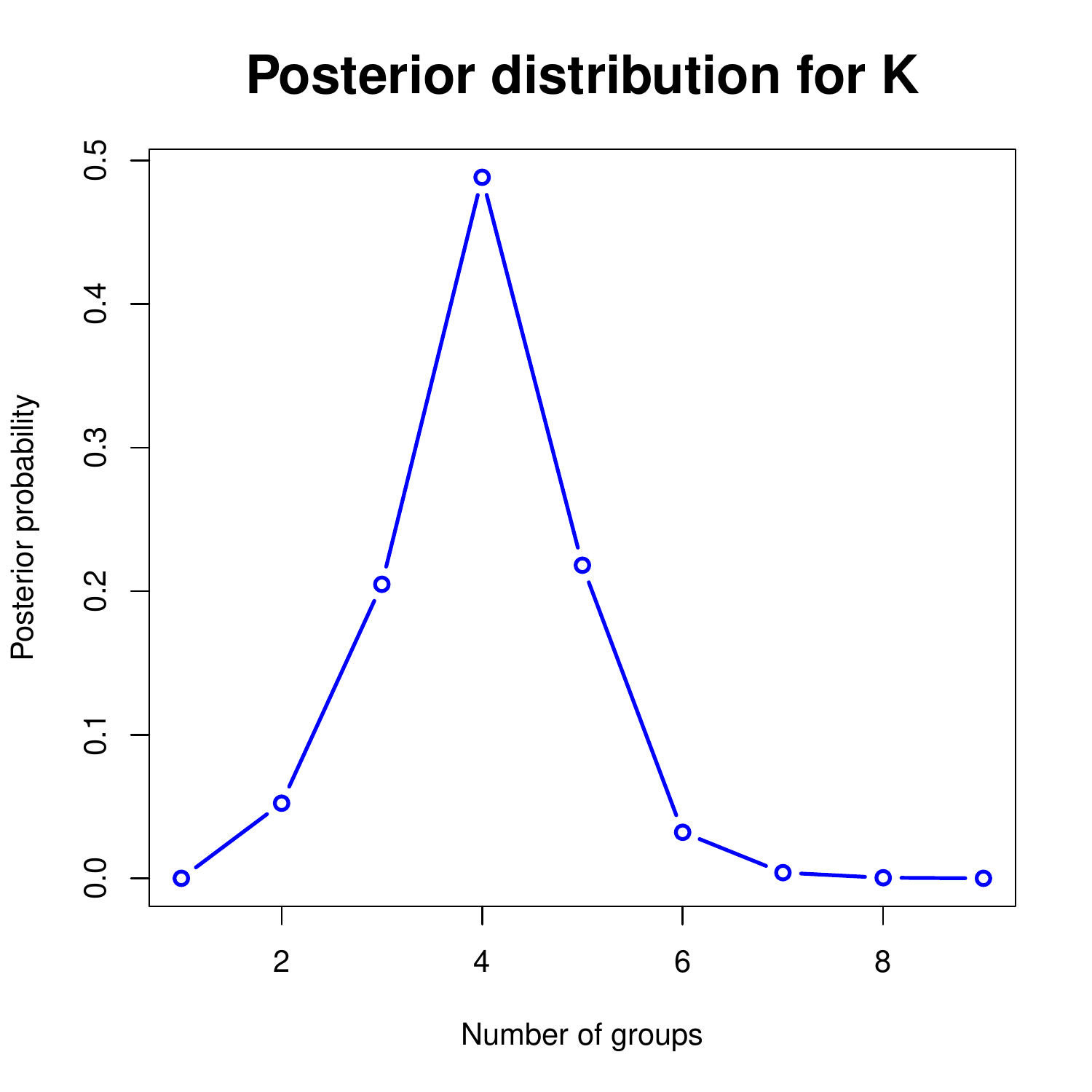}
 \caption{\textbf{Galaxy dataset}. 
 On the left panel, the $VI$ loss best partition found is shown. The right panel shows the posterior probabilities for the number of groups. 
 The distribution has a peak at $K=4$, which contrasts with the number of clusters in the optimal partition, equal to $3$.}
 \label{fig:galaxy1}
\end{figure}

We obtained a sample for the allocations using the collapsed sampler described in Appendix \ref{ASimplifiedAllocationSampler}.
One million observations were first discarded as burn-in, then one observation every hundredth was retained until a sample size of $10{,}000$ was obtained.
The chain appeared to mix well suggesting convergence to the target distribution.
The sample was post-processed using the method described in Section \ref{ClassesOfEquivalencesWithinThePosteriorSample}. 
Then, several runs of the greedy algorithm were performed, using both the noisy MAP and completely random starting partitions. 
The left panel of Figure \ref{fig:galaxy1} shows a histogram of the observed data with the overall best clustering found. 
The number of groups for the VI Bayes action is $3$, which is in line with the results of \textcite{wade2015bayesian}, but notably different from the results of \textcite{richardson1997bayesian}. 

The computational time needed to get the sample was about $45$ seconds, whereas an average of $5$ seconds was required for each run of the greedy algorithm, with $K_{up}$ fixed to $50$ for both algorithms. 

\subsection{Stochastic block models: French political blogosphere}
\subsubsection{The data}
The data first appeared in \textcite{zanghi2008fast} and consist of a undirected graph where nodes represent political blogs' websites and edges represent hyperlinks between them. 
As in \textcite{latouche2011overlapping} we focus only on a subset of the original dataset, available in the R package \texttt{mixer}. 
The data consist of a single day snapshot of political blogs automatically extracted on the 14th of October 2006 and manually classified by the ``Observatoire Pr\'esidentielle project''.
The graph is composed of $196$ nodes and $1432$ edges, and the main political parties are 
the UMP (French ``republican''), UDF (``moderate'' party), liberal party (supporters of economic liberalism) and PS (French ``democrat''), although $11$ different parties appear in total.
The observed data is modelled by the adjacency matrix $\mathcal{Y}$ whose entries are defined as follows:
\begin{equation}
\rowcolors{1}{}{}
 y_{ij} = \begin{cases}
           1 & \mbox{ if an undirected edge between blogs $i$ and $j$ appear; }\\
           0 & \mbox{ otherwise; }
          \end{cases}
\end{equation}
for every $1\leq i < j \leq N$.

\subsubsection{Stochastic block models}
Stochastic block models \parencite{nowicki2001estimation} are finite mixture models for networks, whereby the clustering problem is formulated on the nodes of the network 
and the connection profile of each node is selected by its cluster membership. 
For every $i$, the allocation variable $z_i$ denotes the group to which node $i$ belongs, and, as in the Gaussian finite mixture context, a Multinomial-Dirichlet structure is assumed on these variables.
The number of underlying groups $K$ is unknown and hence to be inferred.
Conditionally on the allocations, the likelihood for the graph $\mathcal{Y} = \left\{ y_{ij}:\ 1\leq i < j \leq N\right\}$ factorises as:
\begin{equation}\label{sbm1}
 P\left( \mathcal{Y}\middle\vert \textbf{z}, \Pi \right) =  \prod_{g=1}^{K}\prod_{h=1}^{K}\prod_{\left\{i: z_i = g\right\}}\prod_{\left\{\substack{j: z_j = h \\ j\neq i}\right\}} \pi_{gh}^{y_{ij}}\left( 1-\pi_{gh} \right)^{1-y_{ij}}.
\end{equation}
Here, $\Pi$ is a symmetric $K\times K$ matrix of connection probabilities, 
where the generic element $\pi_{gh}$ indicates the probability that an edge occurs between a node in group $g$ and a node in group $h$, for any $g$ and $h$ in $\left\{1,\dots,K\right\}$.
Furthermore, each $\pi_{gh}$ is assumed to be a realisation of an independent Beta random variable. The hyperparameters for the Beta and Dirichlet distributions are all set to $0.5$.

Since conjugate priors are used, all of the model parameters can be integrated out analytically. 
It follows that, as in the Gaussian finite mixture context, the quantity $p\left( \textbf{z}\middle\vert \mathcal{Y} \right)$ is available anaylitically and can be targeted by the sampler described in Appendix \ref{ASimplifiedAllocationSampler}.
Further details on the integration can be found in \textcite{mcdaid2013improved,come2015model}.

\subsubsection{Results}
First, we performed block modelling using the variational algorithm implemented in the package \texttt{mixer} and obtained a partitioning to be used as reference.
The optimal variational solution has $12$ groups, which roughly correspond to the political affiliations, as shown in Table \ref{tab:blog_var}.
\begin{table}[htb]
\centering
\footnotesize
\begin{tabular}{ccccccccccccc}
  \specialrule{.1em}{0em}{0em}
  \rowcolor{gray!50}
 & 1 & 2 & 3 & 4 & 5 & 6 & 7 & 8 & 9 & 10 & 11 & 12 \\ 
  \hline
   Cap21 &   2 &   0 &   0 &   0 &   0 &   0 &   0 &   0 &   0 &   0 &   0 &   0 \\ 
   CA &   0 &   0 &   8 &   0 &   0 &   0 &   0 &   0 &   0 &   0 &   0 &   3 \\ 
   FN - MNR - MPF &   4 &   0 &   0 &   0 &   0 &   0 &   0 &   0 &   0 &   0 &   0 &   0 \\ 
   Les Verts &   5 &   0 &   2 &   0 &   0 &   0 &   0 &   0 &   0 &   0 &   0 &   0 \\ 
   PCF - LCR &   5 &   0 &   1 &   0 &   0 &   0 &   0 &   0 &   0 &   0 &   0 &   0 \\ 
   PCF LCR &   1 &   0 &   0 &   0 &   0 &   0 &   0 &   0 &   0 &   0 &   0 &   0 \\ 
   PS &   5 &   0 &   9 &   0 &   0 &   0 &  19 &  18 &   2 &   4 &   0 &   0 \\ 
   PRG &   9 &   0 &   1 &   0 &   0 &   0 &   0 &   1 &   0 &   0 &   0 &   0 \\ 
   UDF &   0 &   1 &   1 &   0 &  24 &   6 &   0 &   0 &   0 &   0 &   0 &   0 \\ 
   UMP &   1 &  24 &   2 &  11 &   2 &   0 &   0 &   0 &   0 &   0 &   0 &   0 \\ 
   liberaux &   1 &   0 &   0 &   0 &   0 &   0 &   0 &   0 &   0 &   0 &  24 &   0 \\ 
   \specialrule{.1em}{0em}{0em}
\end{tabular}
\caption{French blogs: confusion matrix for the variational partition and the political affiliations.} 
\label{tab:blog_var}
\normalsize
\end{table}

Then, we used our methodology to estimate the VI loss optimal partition. 
The sample for the allocation variables was obtained through the Collapsed SBM algorithm of \textcite{mcdaid2013improved}, discarding the first $1$ million updates and keeping
$1$ observation every $100$th thereafter. 
A sample size of $10{,}000$ was then used to perform the greedy optimisation, using both noisy MAP and random starting partitions. 
The computational time needed to get the sample was about $5$ hours, whereas an average of $50$ seconds was required for each run of the greedy algorithm, with $K_{up}$ fixed to $50$. 

\begin{figure}[htb]
\centering
\includegraphics[width=0.49\textwidth]{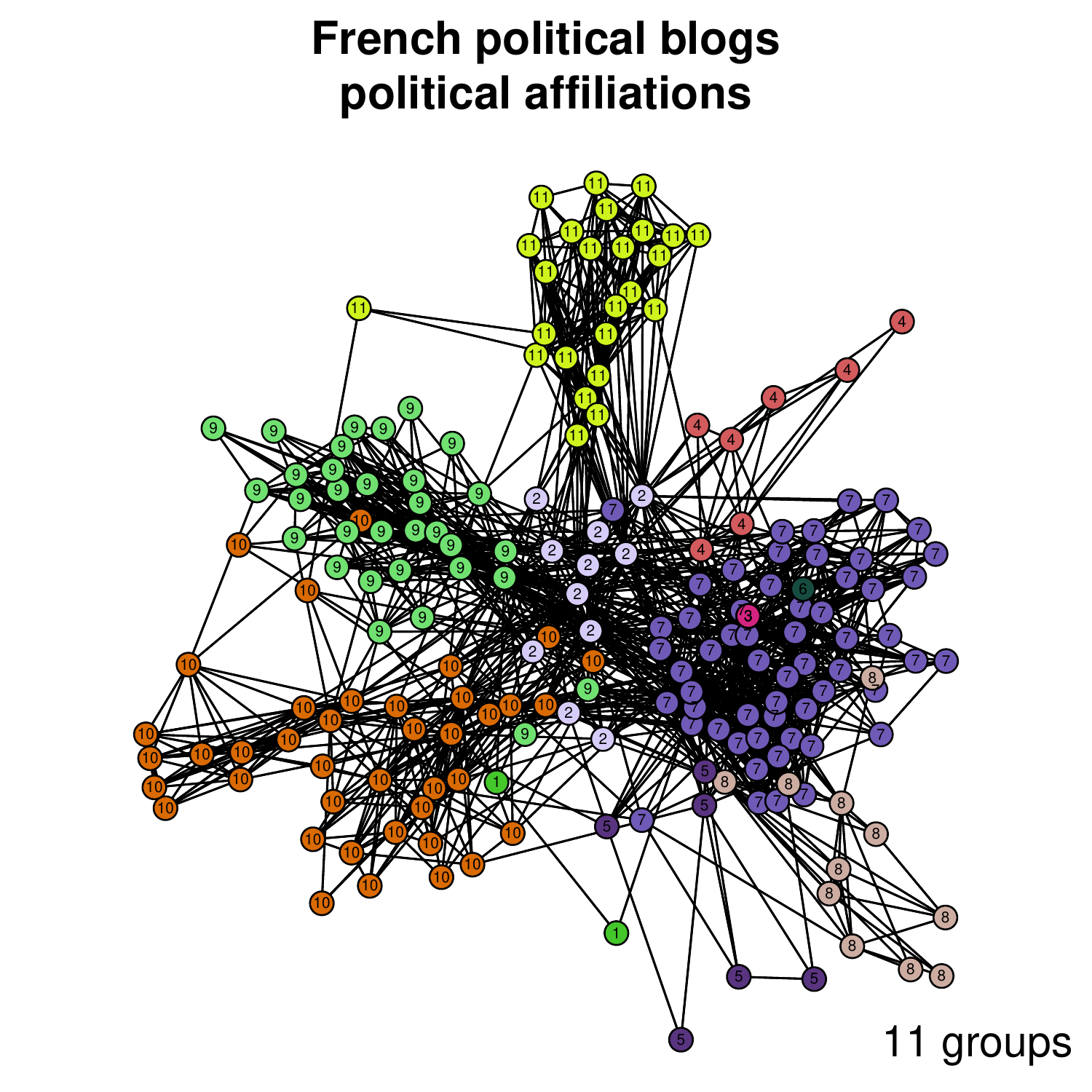}
\includegraphics[width=0.49\textwidth]{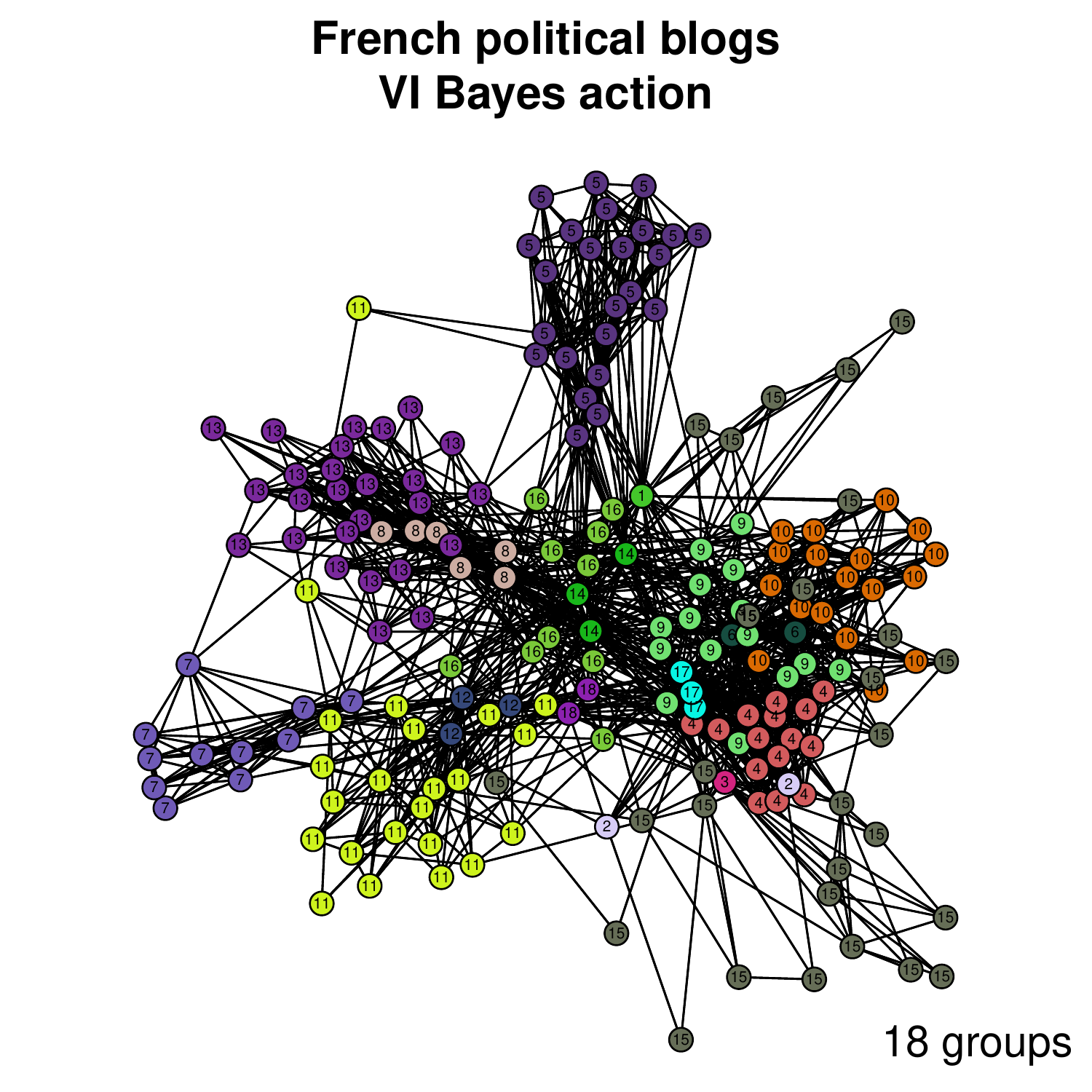}
 \caption{\textbf{French blogs.} Representation of the French political blogs network, with colours and node labels denoting cluster memberships.}
 \label{fig:blog1}
\end{figure}

\begin{figure}[htb]
\centering
\includegraphics[width=0.325\textwidth]{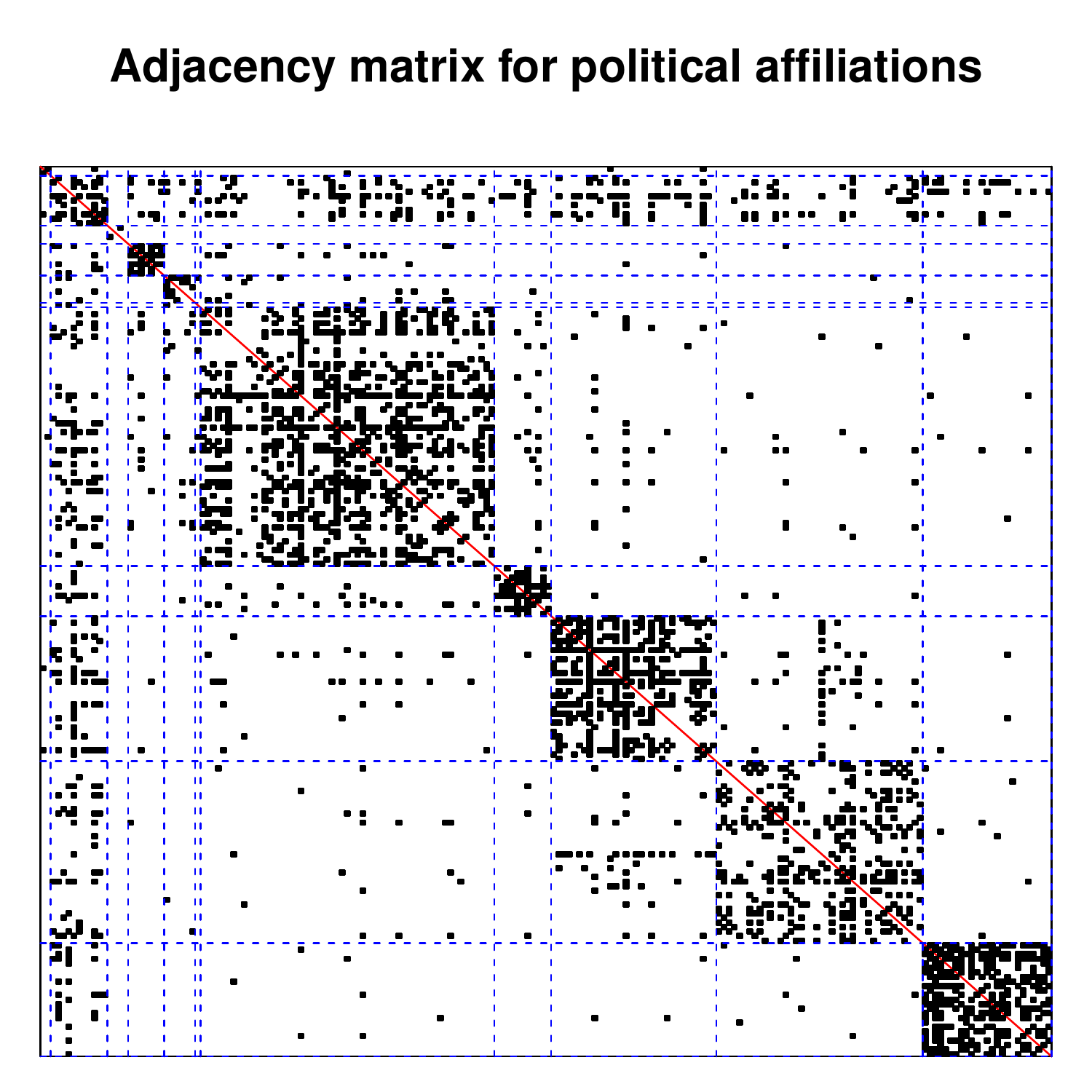}
\includegraphics[width=0.325\textwidth]{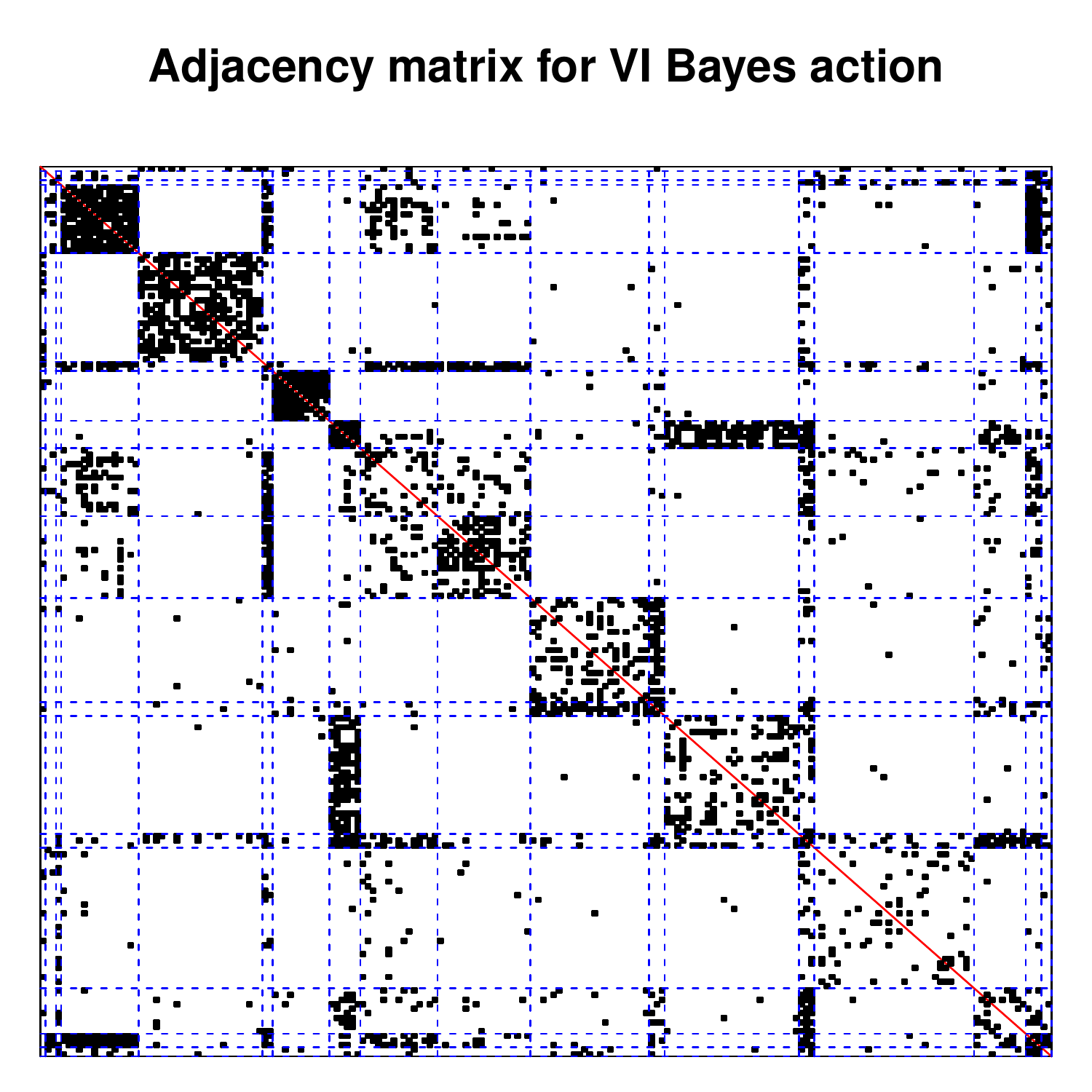}
\includegraphics[width=0.325\textwidth]{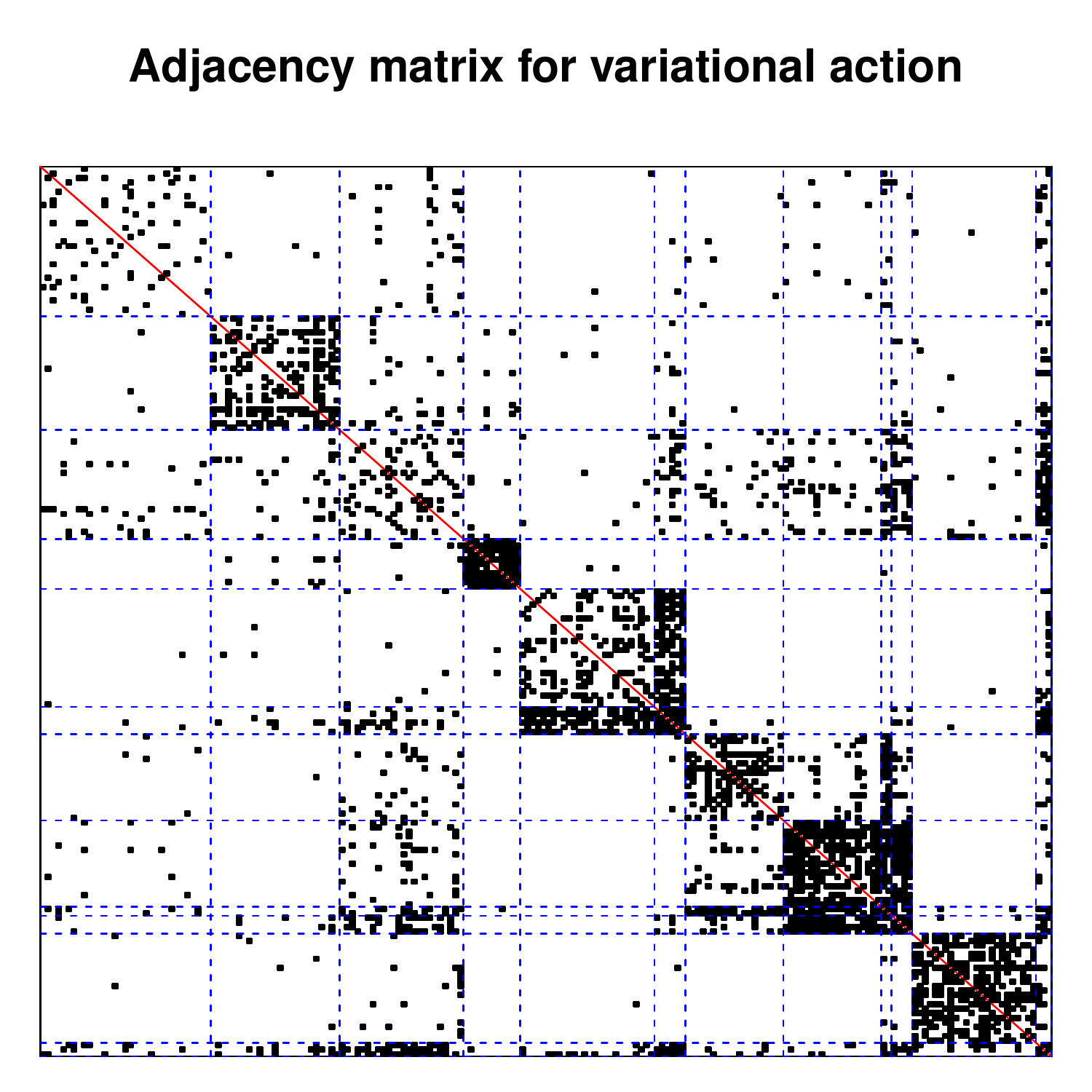}
 \caption{\textbf{French blogs.} Reordered adjacency matrices for three different partitioning of the French political blogs dataset: available political affiliations (left panel),
 VI-optimal allocations (central panel) and variational optimal allocations (right panel).}
 \label{fig:blog2}
\end{figure}

\begin{figure}[htb]
\centering
\includegraphics[width=0.4\textwidth]{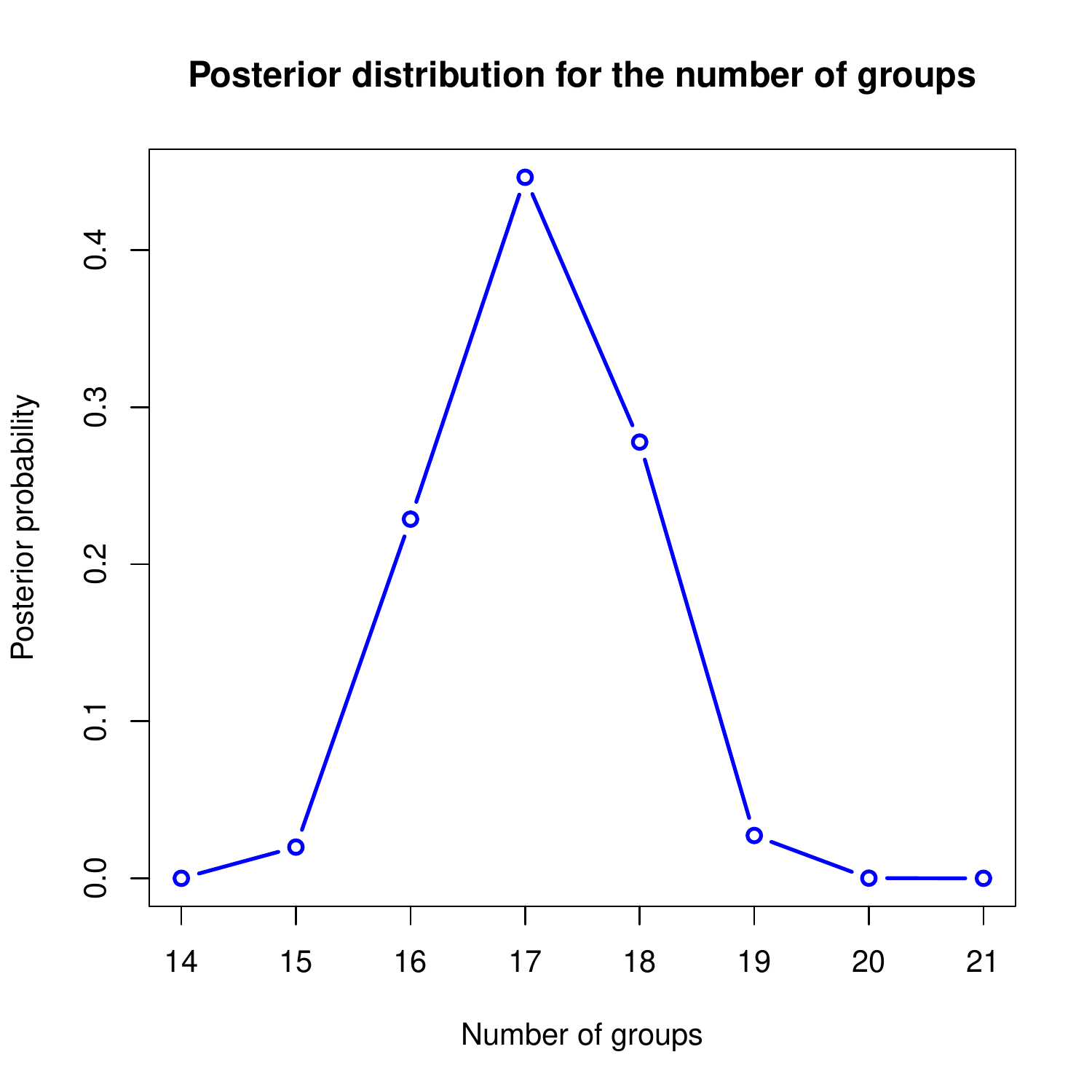}
 \caption{\textbf{French blogs.} Posterior distribution for the number of groups in the French political blogosphere dataset. The MAP value is $K=17$ which contrasts with the optimal value obtained through the greedy algorithm.}
 \label{fig:blog3}
\end{figure}

The VI-optimal partition exhibits $18$ groups, and is represented in the right panel of Figure \ref{fig:blog1}.
Figure \ref{fig:blog2} shows instead the reordered adjacency matrices for the three different partitions. 
The posterior distribution for the number of groups is shown in Figure \ref{fig:blog3}. 
As in the galaxies' dataset, the optimal number of group contrasts with the modal value of the posterior distribution.

It appears that the VI-optimal clustering is a finer partition that splits up some of the larger groups into subgroups. 
Nonetheless from Figure \ref{fig:blog2} it is clear that this entails a better discrimination of the profiles of blogs. 
A confusion matrix matching the solution to the political affiliations is shown in Table \ref{tab:blog_bayes}.
\begin{table}[htb]
\centering
\footnotesize
\begin{tabular}{ccccccccccccccccccc}
  \specialrule{.1em}{0em}{0em}
  \rowcolor{gray!50}
 & 1 & 2 & 3 & 4 & 5 & 6 & 7 & 8 & 9 & 10 & 11 & 12 & 13 & 14 & 15 & 16 & 17 & 18 \\ 
  \hline
   Cap21 &   0 &   0 &   0 &   0 &   0 &   0 &   0 &   0 &   0 &   0 &   0 &   0 &   0 &   0 &   2 &   0 &   0 &   0 \\ 
   CA &   1 &   0 &   0 &   0 &   0 &   0 &   0 &   0 &   0 &   0 &   0 &   0 &   0 &   3 &   0 &   6 &   0 &   1 \\ 
   FN-MNR-MPF &   0 &   0 &   0 &   0 &   0 &   0 &   0 &   0 &   0 &   0 &   0 &   0 &   0 &   0 &   4 &   0 &   0 &   0 \\ 
   Les Verts &   0 &   0 &   0 &   0 &   0 &   0 &   0 &   0 &   2 &   0 &   0 &   0 &   0 &   0 &   5 &   0 &   0 &   0 \\ 
   PCF-LCR &   0 &   1 &   0 &   0 &   0 &   0 &   0 &   0 &   0 &   0 &   0 &   0 &   0 &   0 &   5 &   0 &   0 &   0 \\ 
   PCF LCR &   0 &   0 &   0 &   0 &   0 &   0 &   0 &   0 &   0 &   0 &   0 &   0 &   0 &   0 &   1 &   0 &   0 &   0 \\ 
   PS &   0 &   0 &   0 &  15 &   0 &   2 &   0 &   0 &  13 &  18 &   0 &   0 &   0 &   0 &   5 &   1 &   3 &   0 \\ 
   PRG &   0 &   1 &   1 &   0 &   0 &   0 &   0 &   0 &   0 &   0 &   0 &   0 &   0 &   0 &   9 &   0 &   0 &   0 \\ 
   UDF &   0 &   0 &   0 &   0 &   0 &   0 &   0 &   6 &   0 &   0 &   1 &   0 &  24 &   0 &   0 &   0 &   0 &   1 \\ 
   UMP &   0 &   0 &   0 &   0 &   0 &   0 &  11 &   0 &   0 &   0 &  21 &   3 &   2 &   0 &   0 &   3 &   0 &   0 \\ 
   liberaux &   0 &   0 &   0 &   0 &  24 &   0 &   0 &   0 &   0 &   0 &   1 &   0 &   0 &   0 &   0 &   0 &   0 &   0 \\ 
   \specialrule{.1em}{0em}{0em}
\end{tabular}
\caption{\textbf{French blogs.} Confusion matrix for the VI-optimal partition and the political affiliations.} 
\label{tab:blog_bayes}
\normalsize
\end{table}
The liberals are well discriminated in both the variational and VI-optimal partitions. 
The two partitions also agree on the blogs affiliated to the UDF party: 
$24$ of them are well-discriminated and isolated from the rest, a subset of $6$ blogs are classified into their own group, $1$ blog is associated to the UMP party
and $1$ is not well-recognised.
The main differences between the two partitions arise with respect to the other two relevant parties: UMP and PS. 
In these two cases it appears that the relational profiles of the blogs are not particularly determined by the political affiliation, 
since both partitions recognise a number of subgroups within each party, signaling heterogeneity.
UMP is decomposed in $5$ subgroups in both partitions, while PS is decomposed in $6$ and $7$ subgroups for the variational and VI partition, respectively.

\subsection{Latent block model: Congressional voting data}
We propose an application of our methodology to the UCI Congressional voting data, previously analysed in \textcite{wyse2012block,wyse2014inferring}.

\subsubsection{The data}
The data record whether $435$ members of the $98^{th}$ congress voted ``yay'' or ``nay'' on $16$ key issues. 
Abstained and absent were treated as ``nays''. 
Also, information on the political affiliation of each member is available: $267$ individuals are ``democrats'' and $168$ ``republicans''.
Following \textcite{wyse2012block,wyse2014inferring}, the data are rearranged into a bipartite network, 
whereby two types of nodes are defined (one corresponding to congress members and one to issues) and only undirected edges between nodes of different types are allowed.
Similarly to stochastic block models, an adjacency matrix $\mathcal{Y}$ is used to summarise the data, with edges corresponding to ``yays'' ($y_{ij}=1$) 
and non-edges corresponding to ``nays'' ($y_{ij}=0$).
Note that in this case the matrix $\mathcal{Y}$ has size $435\times 16$, whereby rows correspond to congressmen and columns to issues.

\subsubsection{Bipartite latent block model}
A latent block model (see, for instance, \textcite{wyse2014inferring}) is used to model the bipartite graph.
A clustering problem is formulated on both the rows and columns of the adjacency matrix: two partitions $\textbf{r}$ and $\textbf{c}$ determine the clustering of congress members and issues, respectively.
The number of groups of $\textbf{r}$ and $\textbf{c}$ are denoted by $K_r$ and $K_c$, respectively, and are unknown.
These two partitions independently follow the same Multinomial-Dirichlet structure as described in previous applications.

As concerns the likelihood of the model, a $K_r\times K_c$ matrix $\Pi$ is introduced, so that its generic element $\pi_{gh}\in[0,1]$ corresponds to the probability 
of the occurance of an edge from a node in group $g$ to a node in group $h$. 
Hence, conditionally on the allocations, the likelihood can be factorised into independent blocks:
\begin{equation}\label{lbm1}
 P\left( \mathcal{Y}\middle\vert \textbf{r}, \textbf{c}, \Pi \right) =  \prod_{g=1}^{K_r}\prod_{h=1}^{K_c}\prod_{\left\{i: r_i = g\right\}}\prod_{\left\{\substack{j: c_j = h}\right\}} \pi_{gh}^{y_{ij}}\left( 1-\pi_{gh} \right)^{1-y_{ij}}.
\end{equation}

Bipartite latent block models may also be recast as finite mixture models, where the mixture is with respect to the partitions:
\begin{equation}
 P\left( \mathcal{Y}\middle\vert \boldsymbol{\theta}, \Pi \right) = \sum_{\textbf{r},\textbf{c}} p\left( \textbf{r}\middle\vert\boldsymbol{\theta} \right)p\left( \textbf{c}\middle\vert\boldsymbol{\theta} \right)P\left( \mathcal{Y}\middle\vert \textbf{r}, \textbf{c}, \Pi \right).
\end{equation}

The connection probabilities $\pi_{gh}$ are realisations of independent Beta random variables for every $g=1,\dots,K_r$ and $h=1,\dots,K_c$, and all of the hyperparameters are fixed to $0.5$.

Since conjugate priors are used, all of the model parameters can be integrated out analytically, thereby obtaining the marginal posterior $p\left( \textbf{r},\textbf{c}\middle\vert \mathcal{Y} \right)$ in exact form.
Further details on the integration can be found in \textcite{wyse2012block,wyse2014inferring}.

\subsubsection{Results}
\begin{figure}[htb]
\centering
\includegraphics[width=0.49\textwidth]{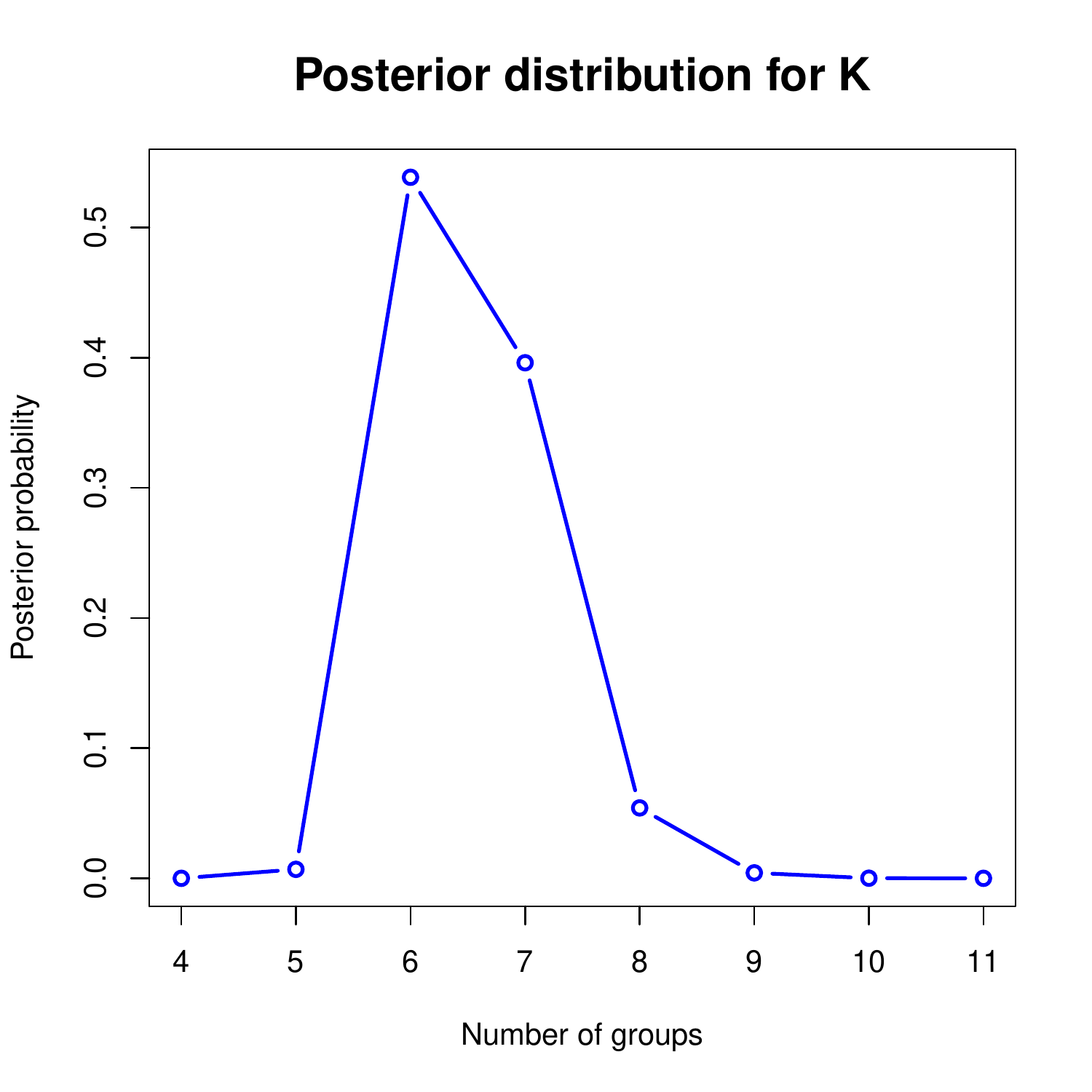}
\caption{\textbf{Congressional voting data}. 
 Posterior distribution for the number of groups of congress members. 
 The VI-optimal value $K=6$ corresponds to the modal value, and the distribution is noticeably right-skewed.}
 \label{fig:congress0}
\end{figure}

\begin{figure}[htb]
\centering
\includegraphics[width=0.3\textwidth]{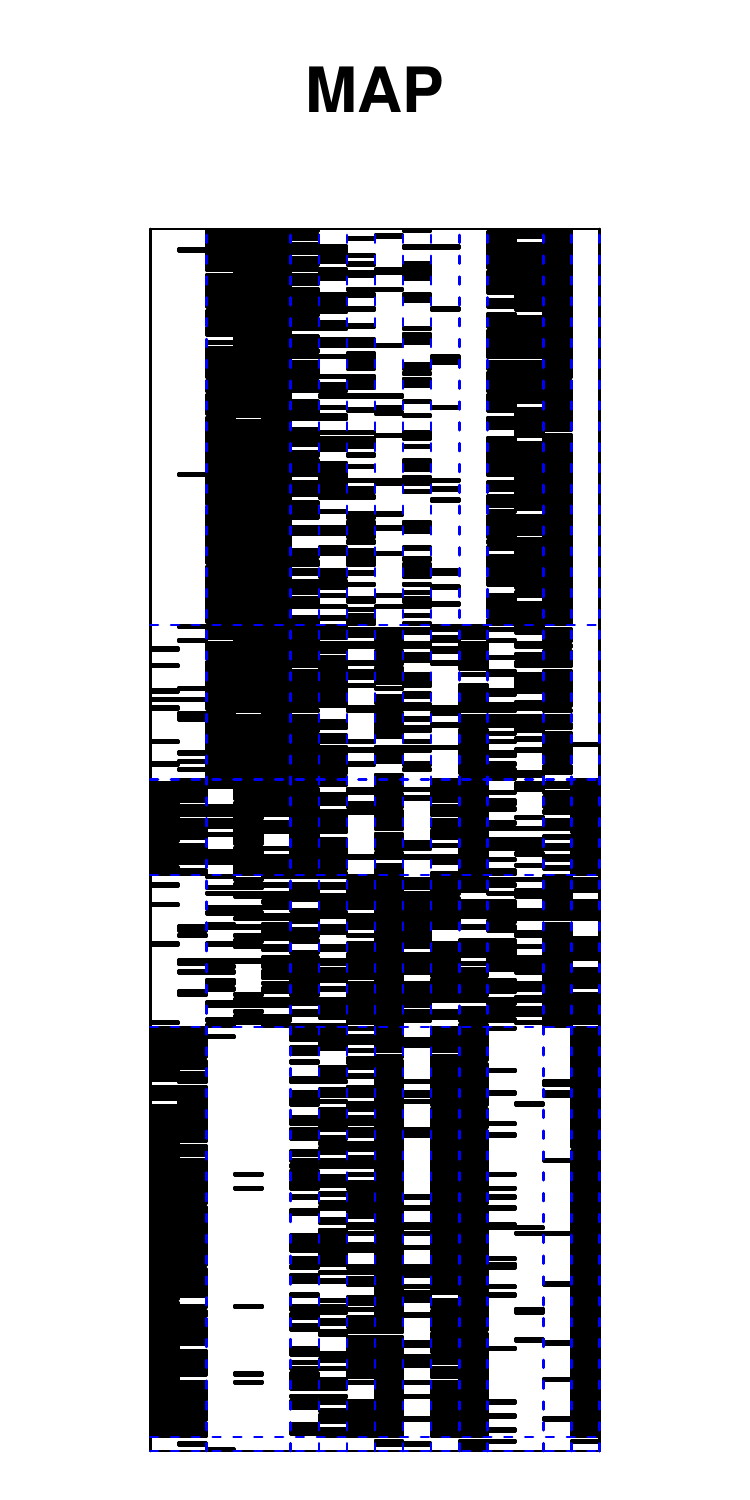}
\includegraphics[width=0.3\textwidth]{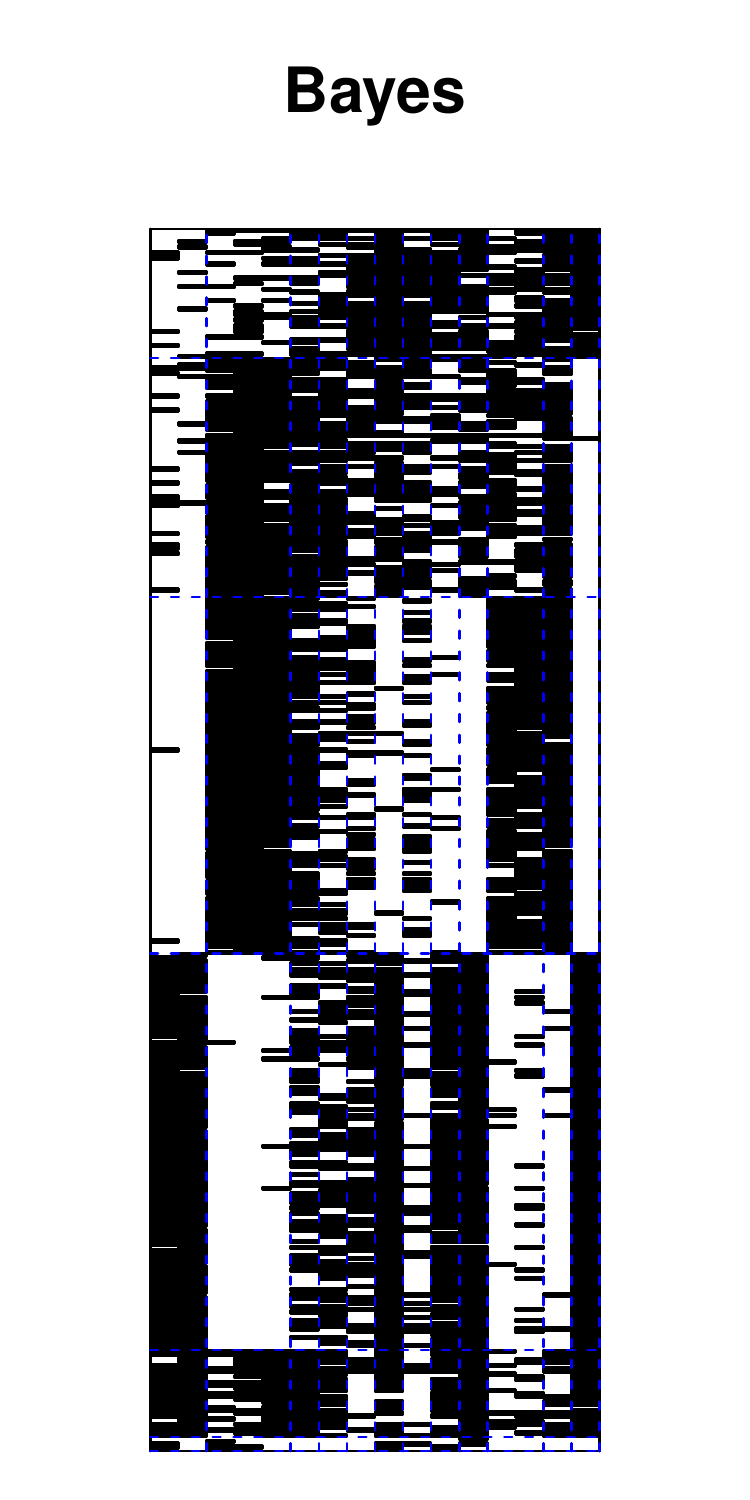}
 \caption{\textbf{Congressional voting data}. Reordered adjacency matrices for the MAP and the VI-optimal partitions. 
 The partitions on the columns (issues) are equivalent, whereas the rows are clustered in different ways, although the number of clusters is equal.}
 \label{fig:congress1}
\end{figure}

The algorithm of \textcite{wyse2012block} was used to obtain a sample for the allocations of both congress members and issues. 
Similarly to previous analyses, $1$ million observations were discarded and $10{,}000$ were used as final sample using a thinning of $100$.
The partitioning of the data corresponding to the highest posterior value was saved as a reference.
We found that posing a clustering problem on the issues was not particularly interesting in that very few issues were aggregated in the same cluster, 
hence we will here show only the cluster analysis on the congress members.
The sample of partitions for the members was processed through the procedure of Section \ref{ClassesOfEquivalencesWithinThePosteriorSample}, and then several runs of the greedy 
optimisation were performed. 
The computational time needed to get the sample was about $30$ hours, whereas an average of $70$ seconds was required for each run of the greedy algorithm, with $K_{up}$ fixed to $30$. 
Figure \ref{fig:congress0} shows the posterior sample for the number of groups.
The reordered adjacency matrices for the MAP and the VI-optimal partition are shown in Figure \ref{fig:congress1}.
From the confusion table shown in Table \ref{tab:congress_table_true_bayes} it appears that the two main political factions are split into $3$ subgroups each,
with a total of $29$ individuals against the tide.
\begin{table}[htb]
\centering
\begin{tabular}{ccccccc}
\rowcolor{gray!50}
  \specialrule{.1em}{0em}{0em}
  & 1 & 2 & 3 & 4 & 5 & 6 \\ 
  \hline
  democrat &  42 &  79 & 127 &  16 &   1 &   2 \\ 
  republican &   4 &   6 &   0 & 125 &  30 &   3 \\ 
  \specialrule{.1em}{0em}{0em}
\end{tabular}
\caption{\textbf{Congressional voting data.} Confusion matrix comparing the political affiliation with the VI-optimal partition.} 
\label{tab:congress_table_true_bayes}
\end{table}

\section{Conclusions}\label{Conclusions}
We have proposed a Bayesian approach to summarise a sample of partitions from an arbitrary clustering context.
We have described a greedy algorithm capable of finding the optimal partition in a wide range of clustering frameworks.
The algorithm can handle many well-known loss functions. 
In our analyses, we focused on the variation of information loss, which has proven to be particularly effective in the optimisation context.

One appealing advantage of our methodology is that it can scale well with the number of items to be classified,
hence being a useful general tool to use in an arbitrary clustering context.
In fact, since previous methods focused only on particular choices of the loss function, our methodology is the only scalable method that can 
encompass most comparison measures within a unified framework.
Also, label-switching issues do not affect our method.

The greedy algorithm usually converges with very few iterations, however several restarts are useful to avoid convergence to local optima. 
We noticed that, when compared to other similar greedy routines \parencite{come2015model,wyse2014inferring,bertoletti2015choosing},
the algorithm is more likely to converge to the global optimum, even though no final hierarchical merge step is used.
This may be a consequence of the fact that the objective function is generally smoother and easier to optimise.

The wide applicability of our algorithm comes at a cost: 
each step of the optimisation process involves a computational cost depending on the size of the sample $T$, which can easily make the problem intractable if a large sample is used.
However, in most cases this impasse can be downsized simply by ``thinning'' the sample.
As concerns storage costs, the the main bottleneck is set by the $T$ contingency tables of size $K^2$ that are used throughout the optimisation.

To emphasise the context-independence of our approach, we have proposed applications to real datasets for three different clustering frameworks.
In the Gaussian finite mixture case (galaxies' dataset), the results look interesting and in line with the work of \textcite{wade2015bayesian}.
The results on the French political blogosphere appear to be very different from those obtained through previous analyses. 
On one hand an overestimation of the number of groups may be argued, on the other the groups obtained with our approach are evidently more homogeneous.
A clustering problem on the members of the congressional voting data has also been proposed: 
here the two main political factions are well recognised and the results seem to agree with the previous analyses of \textcite{wyse2012block,wyse2014inferring}.

For each of the dataset analysed in this paper we have obtained the marginal sample for the allocation variables using a collapsed Gibbs sampler, which is a tool able to explore a number of models at the same time.
This type of approach aims at improving the mixing of the Markov chain while keeping a low computational cost, and it generally works well in many clustering frameworks.
However, due to the discrete nature of the sampled variables, rarely the sampler achieves good acceptance rates, and in some cases this causes a very slow mixing of the chain.
This in turn biases the results obtained through the loss function approach, since our method heavily relies on the good quality of the sample of partitions.
Unfortunately, at the moment there are no good solutions to address this impasse, suggesting that future research should focus on introducing new ways to explore the space of partitions $\mathcal{Z}$ in a clever way, hence making MCMC approaches more efficient.

\section*{Acknowledgements}\label{sec:Acknowledgements}
The authors would like to thank Antonietta Mira for her helpful feedback during the development of the paper.
The authors' research was supported by a Science Foundation Ireland grant: 12/IP/1424.
The Insight Centre for Data Analytics is supported by Science Foundation Ireland under Grant Number SFI/12/RC/2289.

\printbibliography

\newpage

\appendix
\section{Appendix}
\subsection{A simplified allocation sampler}\label{ASimplifiedAllocationSampler}
The methodology described throughout the paper requires a sample of partitions $\textbf{Z}$. 
In practice, such sample may be obtained using Markov Chain Monte Carlo algorithms such as the Reversible Jump algorithm 
\parencite{green1995reversible} and the Allocation Sampler \parencite{nobile2007bayesian}.
In fact, the distinctive feature of these samplers is that they can move across models, hence allowing Bayesian inference on the unknown number of groups. 
The adaptation of these algorithms to an arbitrary clustering context can be quite challenging.
For this reason we describe in this appendix a very simple and general purpose allocation sampler, that can be applied in a wide range of clustering contexts.

The sampler can be thought of as a simplified version of the Allocation Sampler of \textcite{nobile2007bayesian}, where only one type of update step is used.
The marginal posterior for the allocations, denoted by $\pi\left( \textbf{z}\middle\vert \mathcal{Y} \right)$, is assumed to be available in exact form, up to a proportionality constant.
As required by the Metropolis-Hastings algorithm, a proposal distribution $q\left( \textbf{z}'\middle\vert\textbf{z} \right)$ is introduced, denoting the probability of proposing the new 
partition $\textbf{z}'$ when the current partition is $\textbf{z}$. The proposal $q$ is described in the following section.
At each step, the current partition is changed into the new proposed one with probability
\begin{equation}
 \alpha\left( \textbf{z},\textbf{z}' \right) = \min\left\{ 1,\ \frac{q\left( \textbf{z}\middle\vert\textbf{z}' \right)\pi\left( \textbf{z}'\middle\vert \mathcal{Y} \right)}
 {q\left( \textbf{z}'\middle\vert\textbf{z} \right)\pi\left( \textbf{z}\middle\vert \mathcal{Y} \right)} \right\}
\end{equation}
or it is left unchanged otherwise. 
The discrete process so-obtained is a Markov chain over the space $\mathcal{Z}$. 
If the proposal distribution makes such process ergodic, then its stationary distribution corresponds to the marginal posterior distribution
$\pi\left( \textbf{z}\middle\vert \mathcal{Y} \right)$.

\subsection{Proposal distribution}
In \textcite{nobile2007bayesian} the authors introduce a novel proposal distribution specifically designed to sample allocations,
and composed of several types of updates. 
In particular they use the so-called ejection/absorption steps that are meant to enhance the mixing of the chain.
In our sampler we simplify such a proposal distribution essentially confining it to these two steps only.

Given the current partition $\textbf{z}$, let $\underline{N}=\left\{ N_1, \dots, N_{K_{up}}\right\}$ be the vector of counts for the clusters, 
and define the sets $\mathcal{U}=\left\{ g:N_g>0\right\}$ and $\mathcal{E}=\left\{ g:N_g=0\right\}$. $K_{up}$ is the maximum number of groups allowed and may be set equal to $N$. 
The proposal $\textbf{z}'\sim q\left( \cdot\middle\vert\textbf{z} \right)$ is constructed in the following way:
\begin{itemize}
 \item Select an outbound group $g$ uniformly at random in $\mathcal{U}$.
 \item If $|\mathcal{U}| = 1$ or $|\mathcal{U}| = K_{up}$ select an inbound group $h$ uniformly at random in $\mathcal{U}\setminus \{g\}$
 \item Else with probability $0.5$ select $h$ uniformly at random in $\mathcal{U}\setminus \{g\}$ or from $\mathcal{E}$ otherwise.
 \item Once $g$ and $h$ are chosen, set the number of observations $r$ to be moved from $g$ to $h$: $r$ is drawn uniformly at random from 
 $\left\{1,2,\dots,N_g\right\}$ if $N_h>0$ or from $\left\{1,2,\dots,\lceil N_g/2\rceil\right\}$ otherwise.
 \item The items $\mathcal{I}=\left\{ i_1,\dots,i_r\right\}$ that are being moved are chosen uniformly at random within group $g$.
\end{itemize}
The partition $\textbf{z}'$ is then equal to $\textbf{z}$ with the allocations of items in $\mathcal{I}$ changed to $h$.

The probability of proposing $\textbf{z}'$ given $\textbf{z}$ is given by:
\begin{equation}\label{proposal1}
q\left( \textbf{z}'\middle\vert\textbf{z} \right) = 
Pr\left( g \right) Pr\left( h\middle\vert g \right) Pr\left( r\middle\vert g,h \right) Pr\left( \mathcal{I}\middle\vert g,r \right).
\end{equation}
Here follows the explicit formulation of each of the terms on the rhs of \eqref{proposal1}.
\begin{equation}
 Pr\left( g \right) = \frac{1}{|\mathcal{U}|} \mbox{ for all }g\in\mathcal{U}.
\end{equation}
If $|\mathcal{U}| = 1$ or $|\mathcal{U}| = K_{up}$ then 
\begin{equation}
 Pr\left( h\middle\vert g \right) = \frac{1}{K_{up}-1} \mbox{ for all }h\in\mathcal{U}\setminus\{g\},
\end{equation}
otherwise
\begin{equation}
\rowcolors{1}{}{}
 Pr\left( h\middle\vert g \right) = \begin{cases}
                                     \frac{1}{2\left(|\mathcal{U}|-1\right)} \mbox{ for all }h\in\mathcal{U};\\
                                     \frac{1}{2|\mathcal{E}|} \mbox{ for all }h\in\mathcal{E};
                                    \end{cases}
\end{equation}
As concerns $r$:
\begin{equation}
\rowcolors{1}{}{}
 Pr\left( r\middle\vert g,h \right) = \begin{cases}
                                     \frac{1}{N_g} \mbox{ for all }r\in\{1,\dots,N_g\}\mbox{ if $N_h > 0$};\\
                                     \frac{1}{\lceil N_g/2\rceil} \mbox{ for all }r\in\{1,\dots,\lceil N_g/2\rceil\}\mbox{ if $N_h = 0$};
                                    \end{cases}
\end{equation}
and 
\begin{equation}
 Pr\left( \mathcal{I}\middle\vert g,r \right) = \frac{1}{\binom{N_g}{r}}.
\end{equation}

\end{document}